%% file: Main.tex
\documentclass[fleqn,usenatbib]{mnras}

\usepackage{newtxtext,newtxmath}

\usepackage[T1]{fontenc}

\DeclareRobustCommand{\VAN}[3]{#2}
\let\VANthebibliography\thebibliography
\def\thebibliography{\DeclareRobustCommand{\VAN}[3]{##3}\VANthebibliography}

\usepackage{graphicx}	
\usepackage{amsmath}	
\usepackage{longtable}
\usepackage{xspace}

\newcommand{\cd}{d$^{-1}$\xspace}

\title[Detection of frequency groups in $\gamma$\,Cas]{Short-term variability and mass loss in Be stars\\ VI. Frequency groups in $\gamma$\,Cas detected by {\it TESS}}

\author[J. Labadie-Bartz et al.]{
Jonathan Labadie-Bartz,$^{1}$\thanks{E-mail: jbartz@usp.br (JLB)}
Dietrich Baade,$^{2}$\thanks{E-mail: dbaade@eso.org (DB)}
Alex C. Carciofi,$^{1}$
Amanda Rubio,$^{1}$
Thomas Rivinius,$^{3}$
\newauthor
Camilla C. Borre,$^{4}$
Christophe Martayan$^{3}$
and Robert J. Siverd$^{5}$
\\
$^{1}$Instituto de Astronomia, Geof{\' i}sica e Ci{\^e}ncias Atmosf{\'e}ricas, Universidade de S{\~ a}o Paulo, Rua do Mat{\~ a}o 1226, Cidade Universit{\' a}ria, 05508-900 S{\~a}o Paulo, SP, Brazil\\
$^{2}$European Organisation for Astronomical Research in the Southern Hemisphere (ESO), Karl-Schwarzschild-Str.\ 2,
85748 Garching b.\ M\"unchen, Germany\\
$^{3}$European Organisation for Astronomical Research in the Southern Hemisphere (ESO), Casilla 19001, Santiago 19, Chile\\
$^{4}$Stellar Astrophysics Centre, Department of Physics and Astronomy, Aarhus University, Ny Munkegade 120, DK-8000 Aarhus C, Denmark\\
$^{5}$Gemini Observatory/NSF’s NOIRLab, 670 N. A’ohoku Place, Hilo, HI, 96720, USA \\
}

\date{Accepted XXX. Received YYY; in original form ZZZ}

\pubyear{2020}

\begin{document}
\label{firstpage}
\pagerange{\pageref{firstpage}--\pageref{lastpage}}
\maketitle

\begin{abstract}
In photometry of $\gamma$\,Cas (B0.5\,IVe) from the \textit{SMEI} and \textit{BRITE}-Constellation satellites, indications of low-order non-radial pulsation have recently been found, which would establish an important commonality with the class of classical Be stars at large. New photometry with the {\it TESS} satellite has detected three frequency groups near 1.0 ($g1$), 2.4 ($g2$), and 5.1 ($g3$)\,\cd, respectively.
Some individual frequencies are nearly harmonics or combination frequencies but not exactly so.  Frequency groups are known from roughly three quarters of all classical Be stars and also from pulsations of $\beta$\,Cep, SPB, and $\gamma$\,Dor stars and, therefore, firmly establish $\gamma$\,Cas as a non-radial pulsator.  The total power in each frequency group is variable.  An isolated feature exists at 7.57\,\cd and, together with the strongest peaks in the second and third groups ordered by increasing frequency ($g2$ and $g3$), is the only one detected in all three {\it TESS} sectors.  
The former long-term 0.82\,\cd variability would fall into $g1$ and has not returned at a significant level, questioning its attribution to rotational modulation. 
Low-frequency stochastic variability is a dominant feature of the {\it TESS} light curve, possibly caused by internal gravity waves excited at the core-envelope interface. These are known to be efficient at transporting angular momentum outward, and may also drive the oscillations that constitute $g1$ and $g2$.  The hard X-ray flux of $\gamma$\,Cas is the only remaining major property that distinguishes this star from the class of classical Be stars.
\end{abstract}




\section{Introduction}
\label{sec:intro} 
The eponymous emission lines of classical Be stars \citep[for a broad review see][]{2013A&ARv..21...69R} form in a rotationally supported equatorial disk that unequivocally consists of material ejected from the star itself.  Although Be stars rotate closer to the critical rate than all other stars on and near the main sequence, additional energy and angular momentum are required to lift matter into an orbit.  Under the influence of viscous friction \citep{Haubois2012, Rimulo2018} and radiative ablation \citep{Kee2018}, Be disks decay on time-scales of months to years.  Much of the necessary regular replenishment of the disk seems to take place in the form of stellar outbursts.  From space photometry, evidence has been found \citep{2018A&A...610A..70B, 2018pas8.conf...69B, 2020svos.conf...35B} that the superposition of non-radial pulsation (NRP) modes can lead to non-linear coupling, resulting in enhanced amplitudes that are much larger than the sum of the individual mode amplitudes.
In this way, enough energy may be available and released for the ejection of matter into the disk.  A small part of the ejecta accumulates the necessary specific angular momentum through viscous redistribution while most of the rest falls back to the star \citep{2002MNRAS.337..967O,Haubois2012}.  

The largest sample of Be stars with time-series photometry from space is currently being assembled by the {\it TESS} satellite (described in Sect.\,\ref{sec:TESS}).  A first reconnaissance analysis by \citet{2020svos.conf..137L} found a considerable diversity of variability types \citep[see also][]{BalonaOzuyar2020}.  One of them is the presence of distinct frequency groups, here understood as an assembly of frequencies (3 or more) that are usually no more than a few tenths of \cd from each other. They occur in about three quarters of all classical Be stars. Their fraction drops from more than 80\% among early-type Be stars to less than 60\% at late types. However, the decline is also a matter of instrumental sensitivity, as the photometric amplitudes fall sharply towards the late portion of the B spectral type \citep[\textit{e.g.}][]{2007ApJ...654..544S}.
  
Frequency groups in Be stars have also been detected with all other stellar space photometers that have observed Be stars, namely \textit{MOST} \citep{2005ApJ...635L..77W}, \textit{SMEI} \citep{2011MNRAS.411..162G}, \textit{CoRoT} \citep{Semaan2018}, \textit{Kepler} \citep{2015MNRAS.450.3015K}, and \textit{BRITE}-Constellation \citep{2018pas8.conf...69B}.  In the majority of cases, three main groups are found:  one at very low frequencies (hereafter $g0$, $\sim0.05$\,\cd), one centered roughly between 0.5 and 3\,\cd (hereafter $g1$), and the third one (hereafter $g2$) located at approximately twice the frequencies of the middle group and often very crudely twice as wide as the middle one. Additional frequency groups may also exist, typically near integer multiples of $g1$ and $g2$, and usually with relatively lower amplitudes. During outbursts of Be stars, the appearance of frequency groups can change drastically \citep{2009A&A...506...95H}.  

\citet{2017A&A...598A..74P} and \citet{2018pas8.conf...69B} have suggested, and partly illustrated by examples, that, in Be stars, the central group ($g1$) may consist of genuine pulsation frequencies, the low group ($g0$) may be mainly populated by difference frequencies of the middle group, and the high group ($g2$) may be primarily comprised of sum frequencies and first harmonics, also of the middle group.  Frequency groups have also been reported for $\beta$\,Cep, SPB, and $\gamma$\,Dor stars \citep{2012AJ....143..101M} so that they are not specific to Be stars and not even restricted to B-type stars.  Recently, \citet{2018MNRAS.477.2183S} developed the idea that frequency groups form when rapid rotation displaces the eigenfrequencies of sectoral modes such as to coincide - and resonate - with combination and/or harmonic frequencies of the base modes.  

This paper reports the detection of frequency groups in the prominent Be star $\gamma$\,Cas. Sect.\,\ref{sec:gCas_sec} introduces $\gamma$\,Cas and similar stars and discusses the spectroscopic and photometric history of the system up until the epoch of {\it TESS}. The {\it TESS} satellite and its observations, and the extraction of data and their processing are discussed in Sect.\,\ref{sec:obs}. The time-series analysis (TSA) of {\it TESS} data is performed in Sect.\,\ref{sec:TSA}, Sect.\,\ref{sec:discussion} discusses the results 
and places them in context with comparisons to (recent) historical data for $\gamma$\,Cas and other Be stars,
and Sect.\,\ref{sec:conclusions} summarizes the conclusions.

\section{$\gamma$\,Cas}
\label{sec:gCas_sec}

\subsection{Previous observations and $\gamma$\,Cas-like stars}
\label{sec:gCas}
Since $\gamma$\,Cas (27\,Cas, HR\,264, HD\,5394, HIP\,4427, TIC\,51962733; B0.5\,IVe) was the first Be star discovered \citep{1866AN.....68...63S}, it was only natural to consider it the prototypical Be star.  However, \citet{2002ASPC..279..221H} compiled several observations that make the prototype character of $\gamma$\,Cas less obvious but by no means implausible.  $\gamma$\,Cas could be a relatively extreme representative of its cohort, not least because hardly any other Be star has been observed as intensively as $\gamma$\,Cas has.  More challenging to $\gamma$\,Cas's status were (i) the ubiquity of low-order NRPs in Be stars \citep{2003A&A...411..229R, Semaan2018}, which, in $\gamma$\,Cas, remained undetected for a long time \citep{Borre2020}, and (ii) the presence in $\gamma$\,Cas of a thermal X-ray flux at a level intermediate between that from shocks in winds of luminous OB stars and that from accretion in high-mass X-ray binaries but exhibiting a hardness without equivalent in other early-type stars \citep{2016AdSpR..58..782S}.  These X-ray properties are shared by a small but still growing fraction of Be stars \citep{2020MNRAS.493.2511N}.  

The apparent restriction to Be stars of such X-ray properties suggests that the circumstellar disk plays a role in their formation.  \citet{2016AdSpR..58..782S} describe the notion, developed in several earlier papers by the same authors and their collaborators, that the X-rays are due to the interplay between two magnetic fields, one in the disk and one in the photosphere which, however, are not directly detected.  The stellar one was proposed to manifest itself indirectly by rotational modulation of the stellar brightness.  The corresponding 0.82\,\cd frequency was detectable for a decade but disappeared later \citep{HenrySmith, Borre2020}.  The field in the disk was assumed to lead to magneto-rotational instability (MRI) so that gas parcels may fall back to the star, on the way to the photosphere are accelerated by the postulated stellar magnetic field, and finally release their gravito-kinetic energy as X-rays.  Recent observations of the $\gamma$\,Cas-like star $\pi$ Aqr showed that a strong long-term increase in the H\,$\alpha$ line emission was not accompanied by a similar change in the X-ray flux \citep{2019A&A...632A..23N}.  

In the picture of \citet{2016AdSpR..58..782S} for $\gamma$\,Cas, the 0.82\,\cd rotation rate is important for two reasons.  First, it is very nearly the star's critical rotation frequency which \citeauthor{2016AdSpR..58..782S} invoke to activate effective local envelope dynamos.  Second, the highest possible stellar rotation rate is needed to explain the rate of propagation of  migrating subfeatures in spectral lines as corotation of exophotospheric cloudlets with the photosphere.  It should be noted that several of the data strings discussed by \citeauthor{2016AdSpR..58..782S} in this regard hardly cover a single rotation period and even less.  If these features are due to traveling waves such as high-order NRP and possess their own prograde angular velocity, the frequency window of the possible stellar rotation rate obviously becomes much wider.  The model does not state which physical quantity is advected by the rotation and leads to the photometric variability nor does it explain the high phase coherence.  

Near the main sequence, $\sim$10\% of all massive OBA stars possess large-scale magnetic fields \citep[\textit{e.g.}][]{Grunhut2017,Sikora2019}.  The most notable exception to this rule are classical Be stars for which a survey of 85 objects did not yield a single detection \citep[down to a detection limit of $\sim$ 50 -- 100 Gauss;][]{Wade2016}.  This is generally understood as an incompatibility of a stable Keplerian disk with large-scale magnetic fields of $\sim$100 Gauss or stronger \citep{Owocki2006,ud-Doula2018}.  Because the small-scale magnetic fields invoked for X-ray production in $\gamma$\,Cas-like stars (\textit{i.e.} classical Be stars with hard thermal X-ray flux) are too ill constrained by observations, it is not known whether they would lead to a similarly prohibitive magnetic coupling between star and disk.  

Furthermore, the current magnetic concept does not incorporate the presence of a companion star that was discovered in some of these so-called $\gamma$\,Cas-like stars, including $\gamma$\,Cas itself \citep{2012A&A...537A..59N} and $\pi$\,Aqr \citep{2002ApJ...573..812B}.  The companions have not so far been seen directly in any part of the spectrum, but have been inferred from radial velocity measurements of the Be star or its disk. Under the assumption that the secondaries are compact objects, some of the X-ray observations have been attributed to accretion \citep{2016ApJ...832..140H, 2017MNRAS.465L.119P, 2018PASJ...70..109T}.  A more Be-star-specific model was recently developed by \citet{LangerBaade2020} which explains the X-ray flux by the interaction of the fast wind of a helium star with the Be disk. While the \citeauthor{2016AdSpR..58..782S} model predicts that all (early-type) Be stars may develop $\gamma$\,Cas-like X-ray activities, the Be + helium-star and Be + compact object models are obviously only applicable if the assumed companion exists.  Such companions may be numerous but are not easy to detect \citep{WangGiesPeters2018, Klement2019, LangerBaade2020}.  

The first possible evidence of NRP in $\gamma$\,Cas was presented by \citet{NinkovgCaszetOph} and later ascertained by \citet{1988PASP..100..233Y} using additional observations.  It consisted of multiple narrow features traversing photospheric line profiles from the most negative to the most positive velocities.  The inferred azimuthal mode orders, $|m|$, were $\sim12\pm4$.  From a short series of spectra, \citet{HoraguchigCas} confirmed the presence of the line-profile variability a few years later.  While rotational Doppler spreading enhances the spectroscopic visibility of high-order NRP modes, their photometric effects are strongly diluted in integral light.  Low-order modes ($m \approx-2$) that are typical of many other (not quite as hot) Be stars \citep{2003A&A...411..229R, LabadieB2017, Semaan2018} were not reported. Furthermore, ground-based photometry did not indicate multiperiodicity \citep{HenrySmith}, the tell-tale signature of NRP.  Therefore, \citet{Borre2020} analyzed medium-high-cadence photometry by several satellites with a time span of 8+4 years (16 years in total).  The observations confirmed that the 0.82\,\cd variability \citep{2016AdSpR..58..782S} has faded below detectability even from space.  Instead, variability with frequencies of 1.25 and 2.48\,\cd was found, thereby probably re-uniting $\gamma$\,Cas with its non-radially pulsating brethren because both frequencies are too high to be due to any rotational modulation.  However, with just two frequencies, one of which (1.25 \cd) was only marginally detected, the evidence for NRP was not overwhelmingly broad, especially since it could not be confirmed that 2.48\,\cd is associated with the B-star primary and not with, \textit{e.g.}, the rotation of its invisible companion.  

In a recent remarkably thorough study of the $\gamma$\,Cas-like stars $\pi$\,Aqr and BZ\,Cru, \citet{Naze2020} found strong spectroscopic and photometric evidence of high-order prograde tesseral NRP modes ($ l \sim 6 \pm 2, |m| \sim 2^{+2}_{-1}$) with frequencies that fall into the range of p modes ($\sim$ 7 -- 12\, \cd).
These rapid line profile variations in $\pi$\,Aqr were first reported in \citet{2003A&A...411..229R}, and were later estimated to correspond to an $l = -m = 5$ prograde mode in \citet{2005ASPC..337..294P}. 
$\pi$\,Aqr was observed during very different states of the disk but the spectroscopic signature of the migrating subfeatures was unaffected.  Therefore, the variability is stellar, and the range in velocity of these features shows that they are associated with the Be star, not its companion, which is less easily deduced from photometry only.  \citeauthor{Naze2020} do not link the high-order non-radial pulsations to the X-ray properties of $\gamma$\,Cas-like stars but insist that the - in grayscale plots of difference spectra - very similar variability on comparable time-scales to $\gamma$\,Cas itself is the manifestation of corotating exophotospheric cloudlets. 
Interestingly, the authors report that the difference between two frequencies in $\pi$\,Aqr near 11.8\,\cd corresponds to the 3.1\,yr interval between two outbursts with amplitude $\sim$0.25\,mag during the long course of which some pulsation amplitudes changed.  This is similar to what has been described by \citet{2018A&A...610A..70B, 2018pas8.conf...69B} for other Be stars and can involve similarly long time-scales \citep{BaadeNuPup}.  It would be the first such case linked to p mode pulsations.  However, \citeauthor{Naze2020} caution that partial obscuration of the stellar disk by the circumstellar disk may be the cause.  

\citet{BalonaOzuyar2020} have proposed another model with magnetic fields that are too weak for detection with current means.  The model is suggested for Be stars at large and assumes a tilted dipole field that causes matter ejected by reconnection events from magnetic spots to accumulate at two opposite locations in the equatorial plane.  Depending on the relative strength of these two clouds, variations with the stellar rotation frequency or its first harmonic may occur.  Because the trapped matter is in all coordinates only loosely locked and its amount will not be constant, coherent variations are not predicted.  The model may not easily explain the mostly sinusoidal shape of the light curves of many periodic Be stars.

Although \citeauthor{BalonaOzuyar2020} do not point this out, the phenomenon they describe seems to have its spectroscopic counterpart in {\v S}tefl frequencies \citep{1998ASPC..135..348S, 2016A&A...588A..56B}.  Their signature is the variability in anti-phase with respect to each other of double emission peaks in the extreme wings of He\,{\sc i} and other lines and occurs in phases of enhanced mass loss when there is enough gas to produce helium line emission.  {\v S}tefl frequencies are typically $\sim10$\% lower than that of the dominating spectroscopic quadrupole $g$ mode.  Therefore, {\v S}tefl frequencies are thought to form in newly ejected matter orbiting in the inner disk that is not yet equalized in azimuth.  They would, therefore, be unrelated to the actual stellar rotation but to orbital periods in the disk.  However, they would be asymptotic tracers of the critical stellar rotation rate, as it sets an upper limit for the {\v S}telf frequency.  For the same reasons as in the magnetic model of \citeauthor{BalonaOzuyar2020}, phase coherent variability is not expected.  
In spite of the large physical differences between the two competing explanations of the {\v S}tefl frequencies, the expected high similarity of the photometric behavior seems to suggest that photometry alone is not likely to be able to distinguish between them.  However, this is only true if the discussion focuses on just one frequency.  For a more complete understanding of Be stars, the full ensemble of frequencies needs to be considered. 

One important point which plain rotating models necessarily fail to account for is the large-scale, large-amplitude photospheric velocity field that is particularly prominent at low inclinations.  However, this characteristic line-profile variability is well described by non-radial g mode pulsation \citep{2003A&A...411..181M}.

\subsection{Recent spectroscopic and photometric context} \label{sec:spectra}

The \textit{viscous decretion disk} \citep[VDD;][]{Lee1991,Carciofi2011} model successfully describes the majority of observed properties of Be star disks and their variability with time, including the regions in which certain observables are formed and the associated time-scale and magnitude of changes in these observables during epochs of disk build up or dissipation \citep{Haubois2012,Haubois2014}. Of particular relevance to the present study of $\gamma$\,Cas, the visible continuum arises in the inner-most disk (with about 1 -- 2 stellar radii) due to an increased continuum emission by the ejecta, mostly due to hydrogen free-bound recombination and also electron scattering,
and can change rapidly in response to a change in the environment near to the star.  By contrast, the H\,$\alpha$ line is formed over a large region of the disk where viscous time-scales are much longer, and small additions of material (relative to the already large amount of H\,$\alpha$-emitting material) contribute little to the overall line strength.  

The H\,$\alpha$ emission has been gradually and steadily trending upward from about 2001  until the epoch of {\it TESS}, growing in the absolute value of the equivalent width (EW) from $\sim$ 25 {\AA} to 37 {\AA} and in the flux ratio of emission-peak to continuum (E/C) from $\sim$ 3.5 to 5.5 \citep{Pollmann2009,2012A&A...537A..59N,Borre2020}.  
However, a disk has existed for decades even prior to 2001 \citep{DoazangCasV,2002ASPC..279..221H}. This trend has continued beyond the {\it TESS} observations reported in this paper as seen in recent spectra in the Be Star Spectra Database \citep[BeSS\footnote{\url{http://basebe.obspm.fr}};][]{BeSS}. 
In other words, the disk of  $\gamma$\,Cas was strong and apparently growing during the {\it TESS} observations, having been built up over decades, and the amount of material in the disk at the time of the {\it TESS} observations was higher than during the period studied by \citet{Borre2020}. In a strong and dense disk such as this the photometric sensitivity \`a la VDD to additional matter injected may be (partly) saturated when the scattering of stellar photons may increase underproportionally \citep{2018MNRAS.479.2214G}. 
This would (possibly strongly) dilute the photometric signature of any discrete or short-lived episodes of enhanced mass ejection, provided the geometry of the inner disk is not significantly altered. However this effect would not drastically reduce the variability that arises in the photosphere (\textit{e.g.} pulsation).

\section{Observations and data reduction}
\label{sec:obs}

\subsection{The {\it TESS} observatory}
\label{sec:TESS}
The four identical 10\,cm cameras of the Transiting Exoplanet Survey Satellite \citep[{\it TESS};][]{Ricker2015} cover a combined field of view of 24$^{\circ}$ $\times$ 96$^{\circ}$ that extends from the ecliptic plane to one of its poles.  With 13 steps in ecliptic longitude, one hemisphere is observed in one year.  Depending on ecliptic latitude, stars are observed in up to 13 such sectors, each spanning $\sim$27.4\,d in time.  

The focal-plane array of each {\it TESS} camera comprises a 2 $\times$ 2 mosaic of CCDs with 2048 $\times$ 2048 pixels each and records light over the range 600 to 1050\,nm, which includes the conventional $I$ and $Z$ and portions of the $V$, $R$, and $Y$ bands.  
90\% of the flux contained in the point-spread function (PSF) is ensquared within 4 $\times$ 4 pixels, each of which measure 0.35\,arcmin on a side.

Stars brighter than $\sim$7.5 in $I_{\rm C}$ saturate the pixel at the center of the PSF on the {\it TESS} CCDs (but precise light curves are still produced for stars much brighter than this; see Sect.\,\ref{sec:process}). For optimal targets, the noise floor is approximately 60 ppm h$^{-1}$. Photon noise dominates down to $I_{\rm C}$\,$\approx$\,13.5\,mag. Owing to its very wide and elongated orbit, {\it TESS} can perform uninterrupted observations for more than 300\,h, is not plagued by near-Earth radiation belts, and suffers only little from terrestrial and lunar stray light (which is the main source of systematic trends). The Full Frame Images (FFIs) for the entire CCD are available at a 30-minute cadence from which light curves can be extracted for any star observed by {\it TESS}, and a selection of targets are also observed in a 2-minute cadence mode.

\subsection{{\it TESS} observations of $\gamma$\,Cas}
\label{sec:obsdata}
The ranges in Barycentric Julian Date (BJD), the {\it TESS} sectors, the CCDs, and the number of data points used for time-series analysis (TSA) are provided in Table\,\ref{Tab:gCas}. $\gamma$\,Cas was observed in two consecutive sectors (50.3 days; Sectors 17 and 18), followed by a 141-day gap and was again observed in Sector 24 for an additional 26.5 days. This makes it impossible to trace any variations with the binary orbital period of 203.55\,d \citep{2012A&A...537A..59N}.  $\gamma$\,Cas was not among the targets pre-selected for 2-minute cadence observations, so the full frame images with 30-minute cadence, available for download at MAST\footnote{\url{https://mast.stsci.edu/}}, were instead used ({\it TESS} Input Catalog ID: TIC\,51962733).

\begin{table}
\caption{{\it TESS} observation of $\gamma$\,Cas. }
\label{Tab:gCas}
\centering
\begin{tabular}{c c c c c c} 
\hline\hline
BJD start & BJD end & Sector & Camera & CCD & n$_{\rm obs}$ \\
\hline
  2458764.69 &  2458789.69  &  17  & 2 & 1 & 1070 \\
  2458790.67 &  2458815.02  &  18  & 2 & 2 & 1017 \\ 
  2458955.80 &  2458982.26  &  24  & 4 & 4 & 1225 \\
\hline  
\end{tabular}
\end{table}

\subsection{Data extraction and processing}
\label{sec:process}

For each sector, a target pixel file from the {\it TESS} FFIs centered on $\gamma$\,Cas was downloaded with TESScut\footnote{\url{https://mast.stsci.edu/tesscut/}} \citep{Brasseur2019}. The size in pixels was chosen to adequately capture the extent of the bloom columns, which span up to 101 rows (20 $\times$ 130 pixels for Sectors 17 and 18, and 20 $\times$ 150 pixels for Sector 24). Although $\gamma$\,Cas is very bright and highly saturates the CCDs, it is still possible to extract a reliable, high-quality light curve. Two different methods are viable. Since the {\it TESS} CCDs are charge conserving, an aperture mask that encloses all charge-containing pixels is one option. Alternatively, a light curve can be extracted from the PSF wings, excluding saturated pixels and columns (\textit{i.e.} `halo photometry'). Both extraction methods were performed, and are further described in Appendix~\ref{sec:LC-extraction}. The resulting light curves are very similar in that they contain the same features. However, the version extracted using all charge-containing pixels is of higher quality (having higher signal-to-noise and being less affected by systematics such as scattered light) and is adopted for the remainder of this work. Both versions of the light curve are compared in Appendix~\ref{sec:LC-extraction}.

\begin{figure*}
\centering
\includegraphics[width=0.99\textwidth]{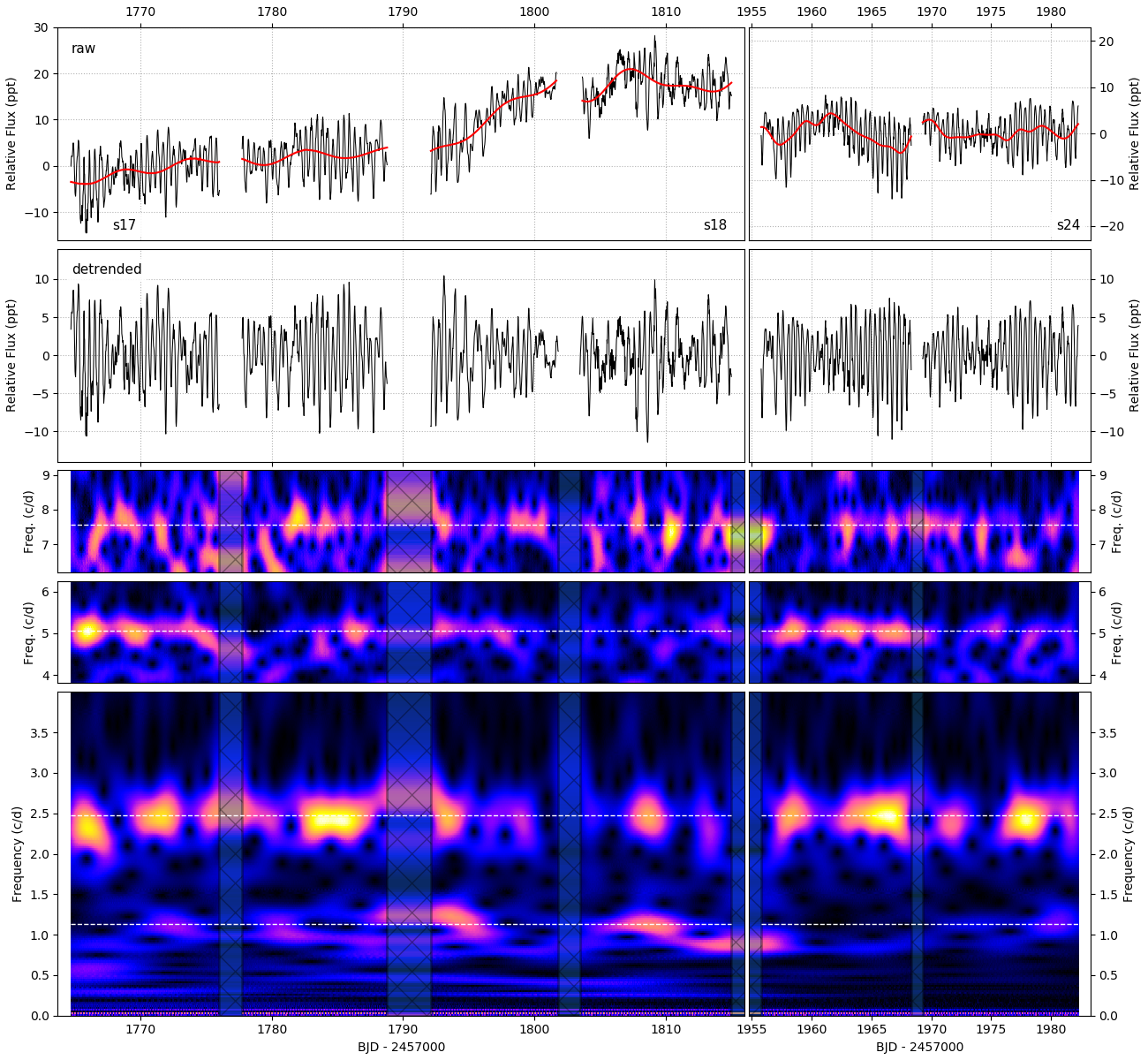}
\caption{\textit{Top:} {\it TESS} light curve of $\gamma$\,Cas showing Sectors 17 and 18 (left) and 24 (right) in black, with low frequency trends traced in red. There is a 141 day gap between Sectors 18 and 24. Sector 18 data are vertically offset from their mean to appear continuous with Sector 17, while Sectors 17 and 24 have their mean value subtracted. \textit{Second from top:} The light curve after removing low frequency trends with time-scales greater than five days. The remaining panels show the wavelet analysis of the detrended light curve. Note that the color scale in the third, fourth, and fifth panels from the top are independent of each other, in order to see the relatively low amplitudes of the two higher frequency groups. However, the color scale is consistent across the full time baseline. Horizontal dashed lines mark the location of the strongest signal in each of the four groups. Hashed regions are gaps in the observations. 
}
\label{fig:wavelet}
\end{figure*}

The resulting light curve is shown in the top panel of Fig.\,\ref{fig:wavelet}. In order to facilitate the extraction of periodic signals, each sector of data was detrended to remove low frequency signals with $f < 0.2$ \cd. The red curve shown in the upper panel of Fig.\,\ref{fig:wavelet} represents these low frequency signals, which are subtracted to produce the detrended version of the light curve that is shown in the second panel of Fig.\,\ref{fig:wavelet}. This detrended version is used to calculate the amplitude spectra and periodic signals discussed throughout this work. Because data for each sector must be extracted separately, it is difficult to determine if trends with time-scales similar to the sector duration ($\sim$25 days) are astrophysical or systematic in nature. This is discussed further in Secs.~\ref{sec:spectra} and~\ref{sec:comparison} and Appendix~\ref{sec:LC-extraction}, especially in regards to the apparent brightening trend in Sector 18.

Within 20\,arcmin from $\gamma$\,Cas, SIMBAD \citep{2000A&AS..143....9W} lists two stars with known magnitude brighter than 10\,mag:  the A5 star BD+59\,148 ($V$\,=\,9.76\,mag, distance: 708") and the F3 star BD+59\,158  ($V$\,=\,9.81\,mag, 738"). Both of these sources are sufficiently distant from $\gamma$\,Cas and the bloom columns caused by saturation that they do not contribute flux to the pixels in the aperture masks used for either version of light curve extraction. This was verified visually for all sectors.

\section{Time-series analysis}
\label{sec:TSA}

A frequency analysis was performed for the entire set of {\it TESS} observations, and for each sector separately. The \textsc{Period04} package \citep{Lenz2005} was used, and the amplitude spectra are shown in Fig.~\ref{fig:multi_freqs} in black.

Because of the short {\it TESS} observing baseline, the window function includes significant side-lobes (Fig.\,\ref{fig:multi_freqs}, inset), which overlap when multiple signals exist at similar frequencies (as is the case with the frequency groups). This can cause spurious peaks in the periodogram with power comparable to real signals. Pre-whitening against each signal is a more reliable approach for determining the underlying frequencies relative to the Lomb-Scargle analysis without pre-whitening. However, some degree of spurious peaks is unavoidable given the nature of variability in $\gamma$\,Cas.  

To determine the individual frequencies present in the data, the \textsc{Vartools} light curve analysis software \citep{Hartman2012} with the \textsc{Lomb-Scargle} routine \citep{Zechmeister2009,Press1992} was used to detect and iteratively pre-whiten the data against each recovered frequency
up to a false alarm probability (FAP) of 10$^{-2}$ \citep[in a similar fashion as][]{2016A&A...593A.106R}, and is also shown in Fig.~\ref{fig:multi_freqs} as vertical red lines (again for the full light curve and each sector separately). The frequency, phase, amplitude, SNR, FAP and group membership are printed in Table~\ref{tbl:freqtbl}, with the phase being measured relative to the time of the first observation of the data analyzed (BJD = 2458764.69 for Sector 17 and the full light curve, 2458790.67 for Sector 18, and 2458955.80 for Sector 24).  The noise floor is frequency dependent (\textit{i.e.} the frequency spectrum has a red noise component), and a threshold of four times the noise level (further discussed and defined in Sec.~\ref{sec:stochastic}) is used to determine the signals reported in Table~\ref{tbl:freqtbl}.

Both methods, with and without pre-whitening, are generally consistent in their identification of the dominant signals. Whenever specific values for frequency or amplitude are used in the following, they are from the pre-whitening analysis of the combined light curve extracted from all count-containing pixels unless otherwise stated. Formal error estimates are given for the frequency values, according to the method of \citet{Montgomery1999}, which assumes Gaussian uncorrelated noise. These, however, represent lower limits for the following reasons. All but the highest frequency signals detected in each light curve exist in densely populated groups, where there is significant overlap between the window function of adjacent signals, adding some degree of additional uncertainty in the values of the recovered frequency and amplitude as first explored in \citet{1978Ap&SS..56..285L}. The variable amplitude of the groups further complicates this. In terms of error estimation, the frequency groups, although astrophysically real, can be considered as a form of correlated noise, 
meaning that error estimates computed analytically under the assumption of Gaussian uncorrelated noise will be too low. 
For these reasons, in general the most relevant quantities here are not necessarily the frequency and amplitude of specific frequencies, but more so the location, strength, width, structure, and variability of the groups themselves. The most notable exceptions to this are the strongest signal at $f = 2.4796(1)$ \cd, and the isolated high frequency signal at $f = 7.571(1)$ \cd, which are introduced below.

\subsection{Frequency groups} \label{sec:freq_groups}
There are two main frequency groups. The first (group 1 = $g1$) is centered around 0.998 \cd (the strongest signal being at $f = 1.1346(6)$ \cd) and the second group, $g2$, is centered around 2.38 \cd (with its strongest signal at $f = 2.4796(1)$ \cd). A third group ($g3$) centered near 5.07 \cd is also present (the strongest signal being at $f = 5.0559(5)$ \cd), as is an isolated, single signal at $f = 7.571(1)$ \cd ($g4$)\footnote{Although only one signal is detected in this `group,' for the sake of consistency in nomenclature this is hereafter referred to as $g4$.}. These groups were identified visually from the amplitude spectrum, and are consistent with the dominant peaks identified through standard frequency analysis methods (\textit{i.e.} those reported in Table~\ref{tbl:freqtbl}). The centers of $g1$, $g2$, and $g3$ reported above are the average of the frequencies (weighted by amplitude) reported for the `All Sectors' light curve listed in Table~\ref{tbl:freqtbl}.

$g3$ appears much narrower than $g1$ and $g2$ (although this could be a matter of instrumental sensitivity), with a width of approximately 0.04 \cd (taken to be the difference between the highest and lowest frequency signals for the full light curves shown in Table~\ref{tbl:freqtbl}), compared to widths of $\sim$0.3 \cd for $g1$ and $\sim$0.7 \cd for $g2$. Note, however, that this convention likely underestimates group widths, as the wings at the edges of the groups (especially in $g1$ and $g2$) are not accounted for. There are no obvious harmonic relationships between the strongest signals of these groups. $g2$ is centered around slightly more than twice the frequency of $g1$, and $g3$ is slightly higher than twice the frequency of $g2$. The signal at 7.571(1) \cd likewise is not obviously harmonically related to any other frequencies, but is 1.4975(2) times the frequency of the signal at $f = 5.0559(5)$ \cd, and is very close to the sum of the strongest signals in $g2$ and $g3$, differing by 0.036(1) \cd. It is however possible for more complex combination frequencies to manifest in the power spectrum of pulsators, including classical Be stars \citep[\textit{e.g.}][]{2015MNRAS.450.3015K}. Partly owing to a dearth of well-defined frequencies that stand out in $g1$ and $g2$, no such attempt was made to robustly test combination frequencies in the {\it TESS} data for $\gamma$\,Cas. 
There are no signals detected at frequencies higher than 7.6 \cd; with the 30-minute cadence, the Nyquist frequency is 24\,\cd.

\begin{figure*}
\centering
\includegraphics[width=0.98\textwidth]{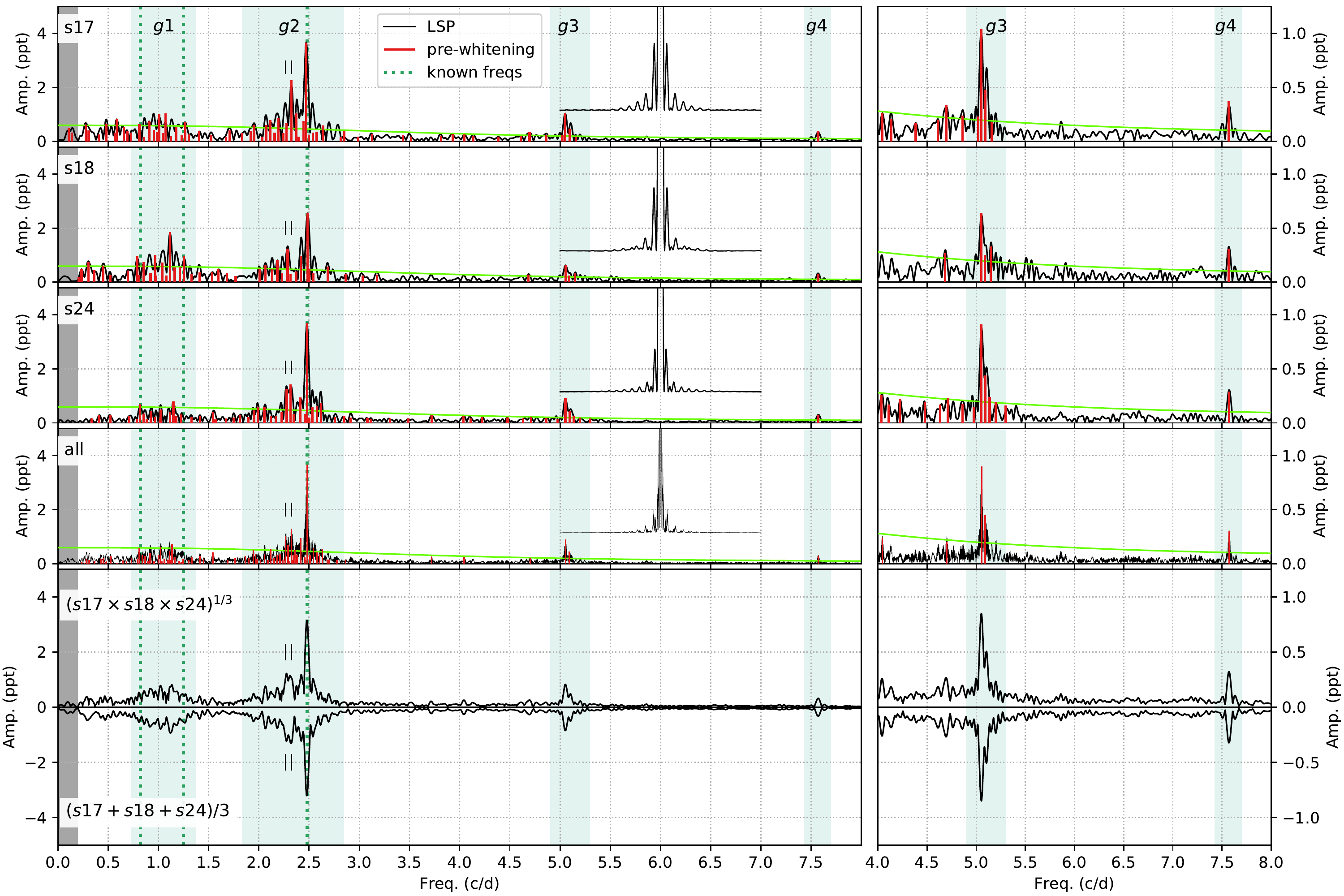}
\caption{A representation of the power spectrum computed from a classical Lomb-Scargle treatment (black) and through applying the Lomb-Scargle technique with iterative pre-whitening (red vertical lines). The dotted green vertical lines mark the historical frequencies at 0.82 \cd, 1.25 \cd, and 2.48 \cd, and the two short vertical black lines mark the location of the second and third strongest signals in $g2$. The blue shaded regions indicate the frequency groups (but do not quantitatively reflect their widths as defined in Sec.~\ref{sec:freq_groups}). The green curve shows the detection threshold used for the signals given in Table~\ref{tbl:freqtbl}, as defined in Sections~\ref{sec:freq_groups} and ~\ref{sec:stochastic}. From top to bottom, the panels show Sectors 17, 18, 24 and the full light curve, respectively. The lower-most panel presents the multiplication and mean (with inverted amplitude) of the periodograms from Sectors 17, 18 and 24. The window function for each is centered at 6 \cd, and with an arbitrary vertical scale. There are no signals detected beyond 8 \cd. The grey bar from 0 to 0.2 \cd identifies the region where low frequency trends were removed prior to the frequency analysis. 
}
\label{fig:multi_freqs}
\end{figure*}

In addition to the clearly dominant signal of 2.4796(1) \cd, there are two other frequencies in $g2$ that seem to stand out, these being the second and third highest amplitude signals detected in the amplitude spectrum for the full light curve (see Table~\ref{tbl:freqtbl}). While these signals are closely spaced, differing by 0.0581(4) \cd, and are not totally resolved in the individual sectors, each sector has signals that appear to correspond to one or both of these frequencies (and they are also present in the sum and product of the amplitude spectra of all 3 sectors). These two signals are marked with short black vertical lines in Fig.~\ref{fig:multi_freqs}.

Because of the short observational baselines, and the fact that data reduction must be done separately for each sector, there is significant difficulty in measuring signals with periods greater than $\sim$10 days in {\it TESS} data. Other studies find a low frequency group ($g0$) in a large fraction of Be stars, typically located around 0.05 \cd (see Sec.~\ref{sec:intro}), which can neither be confirmed nor ruled out in the present analysis of $\gamma$\,Cas. Likewise, variability associated with the orbital period of 203.55 d for $\gamma$\,Cas is inaccessible with {\it TESS}.  However, the overall brightening trends seen in Sectors 17 and 18 (Fig.~\ref{fig:wavelet}) may be real (see Sec.~\ref{sec:comparison}).

\subsection{Group variability} \label{sec:group_var}

The position of the groups does not vary significantly between Sectors 17, 18, and 24, but their strength and/or their composite frequencies do (with the exception of $g4$). The most notable aspects of this variation are the increase in strength of $g1$ from Sector 17 to 18, followed by a decrease between Sectors 18 and 24, and the decrease in strength of $g2$ from Sector 17 to Sector 18 followed by a return to near its original strength in Sector 24. The most significant signal in $g3$ (near 5 \cd) decreases in amplitude by $\sim$40\% from Sector 17 to 18, and also increases back towards its original strength in Sector 24. The signal near 7.6 \cd ($g4$) may decrease in amplitude slightly after Sector 17, but this is less certain and of a lower level compared to the variability of the other groups. 

In an attempt to coarsely visualize where the variability does or does not appear coherent, the lower panel of Fig.\,\ref{fig:multi_freqs} compares the product of the amplitude spectra of Sectors 17, 18, and 24 to their sum. The product emphasizes Lomb-Scargle features that are persistent in time, \textit{i.e.}, are strong at all times and do not significantly shift in frequency. This view of the amplitude spectrum illustrates a few aspects of the data. Especially in regards to $g1$ and $g2$, the densely populated groups and relatively wide window function lead to a situation where the amplitude spectrum within a group never approaches zero. In $g1$ there are no signals that clearly stand out above the `continuum' of the group. In $g2$, the three strongest signals are apparent in both the sum and product amplitude spectrum. Likewise, in $g3$ and $g4$, the structure appears the same in the sum and product versions, suggesting stability of the frequencies (but not necessarily the amplitudes) of these signals.

A wavelet analysis was performed to determine how the power of the detected frequencies varies with a higher temporal resolution compared to simply computing the power spectrum separately for the different sectors. The Python package \textsc{scaleogram}\footnote{\url{https://github.com/alsauve/scaleogram}} was used for this purpose. The wavelet analysis is shown in Fig.~\ref{fig:wavelet}. The variability in $g1$, $g2$, and $g3$ is apparent. However, the structure of the variable group power has some complexity. With many frequencies in $g2$, a complex beating pattern in the light curve can manifest in the wavelet plot as a variable group amplitude without a well defined period. The same logic may apply to the variable amplitude of $g1$, however there is an absence of any coherent dominant signals. The structure of $g3$ is relatively simple. In the amplitude spectrum of the full light curve, $g3$ is made up of two signals spaced by 0.0351(1) \cd. The isolated signal at $f = 7.571(1)$ \cd is generally constant from sector to sector, but appears mildly variable in strength on shorter time-scales. This may simply be an artifact owing to its low amplitude, or any even lower amplitude signals below the detection threshold in the vicinity.

\subsection{Stochastic variability} \label{sec:stochastic}

\begin{figure}
\centering
\includegraphics[width=0.475\textwidth]{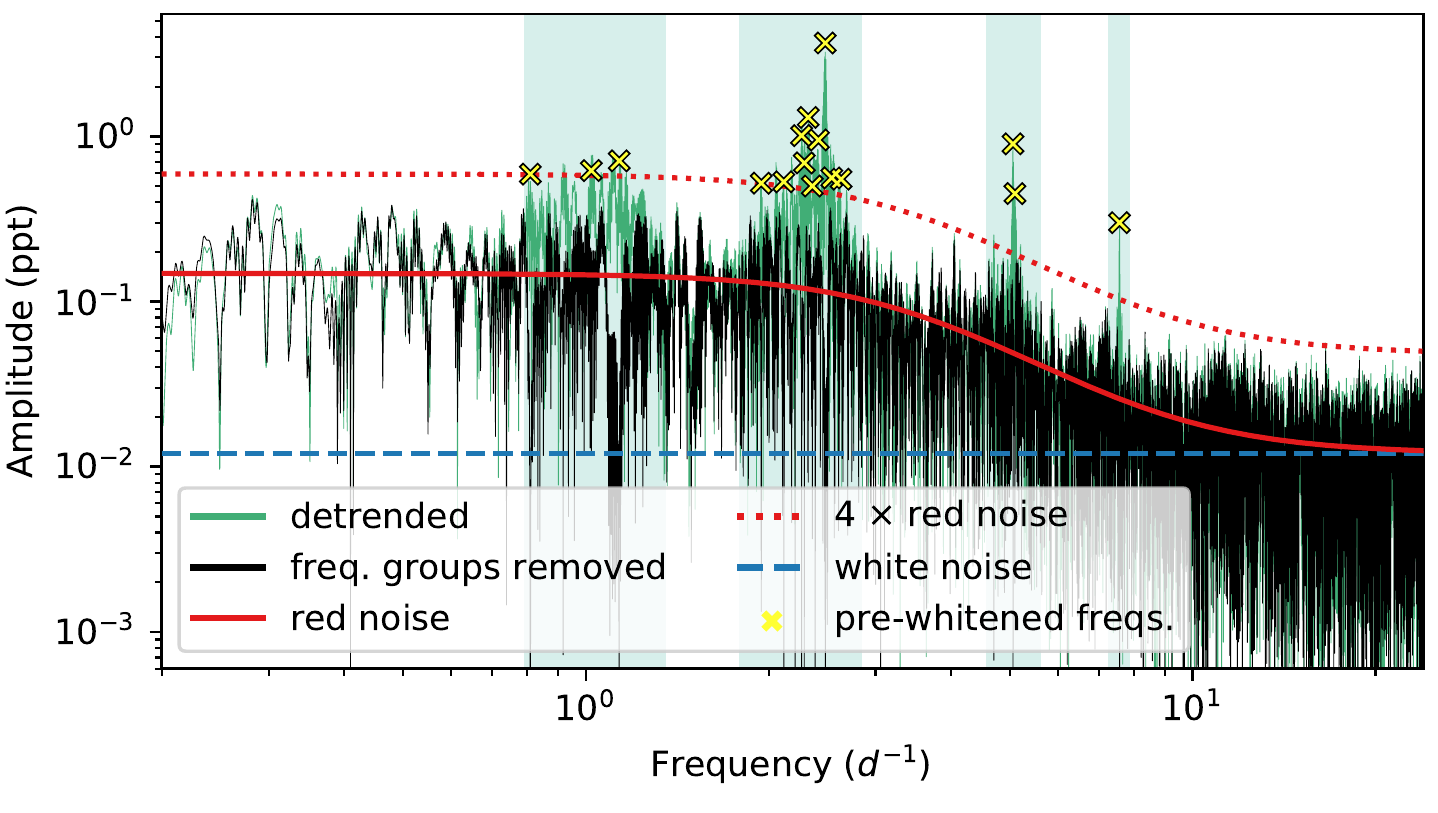}
\caption{Amplitude spectrum of $\gamma$\, Cas. The green curve is identical to what is plotted in Fig.~\ref{fig:multi_freqs} for the combined light curve, but on a log-log scale. The black curve is computed in the same way, but after pre-whitening against the signals in the frequency groups, which is then fit with Equation~\ref{eq:1} (solid red line). The white noise floor is marked by a horizontal dashed line. The red noise profile is then multiplied by a factor of 4 (dotted red line), which is used as the threshold above which the periodic signals identified through iterative pre-whitening are considered to be significant (which are given in Table~\ref{tbl:freqtbl}). Shaded rectangles approximately mark the frequency groups.}
\label{fig:noise_spectrum}
\end{figure}

In a study of 70 OB stars observed with {\it TESS}, each member of the sample exhibits a certain level of stochastic variability that manifests as red noise in their photometric amplitude spectra \citep{Bowman2020}. $\gamma$\,Cas is no exception to this rule. Fig.~\ref{fig:noise_spectrum} shows the amplitude spectrum of $\gamma$\,Cas on a log-log scale, along with the red noise profile as defined below. The red noise profile is then multiplied by 4 which is used as a threshold\footnote{This threshold, determined from the combined light curve, is also applied to each individual sector.} for the signals reported in Table~\ref{tbl:freqtbl}, above which the signals identified through pre-whitening are shown. The morphology of the amplitude spectrum carries information about the physical origin of the stochastic variation. Following the prescription of \citet{2019NatAs...3..760B,Bowman2019,Bowman2020}, in order to parameterize this signal for convenient comparison to recent models, a functional fit of the form \citep[][their equation 2]{Bowman2020}
\begin{equation} \label{eq:1}
\alpha(\nu) = \frac{\alpha_{0}}{1 + (\frac{\nu}{\nu_{char}})^{\gamma}} + C_{w} ,
\end{equation}
is applied, where $\alpha_{0} = 0.136$ ppt is the amplitude at $\nu = 0$ \cd, $\gamma = 2.958$ is the logarithmic amplitude gradient, $\nu_{char} = 3.618$ \cd is the characteristic frequency of stochastic variability, and $C_{w} = 0.012$ ppt is the constant white noise component. The above function with these values is shown as the solid red line in Fig. ~\ref{fig:noise_spectrum}. These values for these parameters are the result of a Markov chain Monte Carlo (MCMC) fit following the method of \citet{Bowman2020}. 
\citet{2020MNRAS.498.3171N} 
fitted the same function but deduced fairly different parameter values.  Several circumstances probably contributed to the difference:  \citet{2020MNRAS.498.3171N} 
applied a much simpler signal-extraction method, which clips much of the valid photometric signal (see Appendix\,\ref{sec:LC-extraction}), used Sectors 17 and 18 only (\textit{i.e.}, without Sector 24), joining them in a different fashion (in their light curve, there is a large jump between the two sectors), and apparently did not remove trends and frequency groups prior to the red-noise measurements (only the strongest signal each in $g2$ and $g3$ were removed). The consequences of this are their higher values for $\alpha_{0}$ (0.31 ppt) and $C_{w}$ (0.025 ppt), and lower values for $\gamma$ (1.82) and $\nu_{char}$ (1.15 \cd).

The parameters derived here for $\gamma$\,Cas are typical for the sample of \citet{Bowman2020}, where it was determined that internal gravity waves excited by core convection are the most likely physical explanation for this feature given the values and ranges of $\alpha_{0}$, $\gamma$, and $\nu_{char}$ for the sample \citep{Rogers2013,Edelmann2019,Horst2020,Ratnasingam2020}. The red noise characteristics of $\gamma$\,Cas also lend themselves to the same conclusion, since internal gravity waves are predicted to have values of $0.8 \, \leq \gamma \, \leq \, 3$, and still contribute power to the frequency spectrum even out towards 100 \cd. Although \citet{Bowman2020} prefer the internal gravity wave explanation, it remains plausible that sub-surface convection zones can produce similar observables in OB stars \citep[][although it seems difficult for sub-surface convection zones to produce an amplitude excess out to the observed relatively high frequencies]{Lecoanet2019}. Furthermore, there are no classical Be stars among the 15 B dwarf stars in the \citet{Bowman2020} sample, and the mean $v\, sin \, i$ value of the 15 B dwarfs is only 40 km s$^{-1}$ (with a maximum of 162 km s$^{-1}$) -- far lower than typical values of the near-critically rotating Be stars. It is therefore not a foregone conclusion that internal gravity waves are a ubiquitous feature of classical Be stars, as they appear to be for the more modestly rotating collection of \citet{Bowman2020}.

\input{freq_tbl_3.tex}

\section{Discussion}
\label{sec:discussion}

Besides the stochastic variability discussed in Sec.~\ref{sec:stochastic} and emphasized in Fig.~\ref{fig:noise_spectrum}, the frequency groups $g1$ and $g2$ are the dominant features in the {\it TESS} light curve of $\gamma$\,Cas. These complexes are much wider than the core of the window function (Fig.\,\ref{fig:multi_freqs}) and, therefore, unambiguously identify $\gamma$\,Cas as a multi-mode non-radial pulsator (since it is not plausible that all of the periodic photometric signals are circumstellar or directly related to rotation considering the time-scales and lack of exact harmonics). The visualisations of the variability as two sets of many individual frequencies (Fig.\,\ref{fig:multi_freqs}) with strong mutual beat patterns in light curves (Fig.\,\ref{fig:wavelet}, top panel) and as wavelet transforms of four frequency ranges with strong amplitude variations (Fig.\,\ref{fig:wavelet}, bottom panels) are complementary presentations of the same behavior \citep{2016A&A...593A.106R}.  The considerable variability in the total power of each frequency group revealed by Fig.\,\ref{fig:wavelet} (bottom panels) is matched by changes in the appearance of the light curve (top panels).  Because this activity does not seem to be periodic (although a time-scale of $\sim5$\,d may be indicated for $g2$), more than two frequencies must be involved, provided that the frequencies are constant. 

Considering the wavelet analysis in Fig.~\ref{fig:wavelet}, the variations in power of the groups are not globally correlated.
Although $g3$ and $g4$ appear at first glance to be qualitatively different from $g1$ and $g2$, the former pair being much narrower and relatively simple, this could just be a matter of reduced photometric amplitude with increasing frequency, and the noise floor of the data. Although $g3$ only has two confidently detected frequencies in the framework of the amplitude spectrum of the full light curve, its variability seen in the wavelet analysis of Fig.~\ref{fig:wavelet} suggests a higher degree of complexity. If $g3$ only contained these two frequencies, a beat period of 28.5 days is expected. While this seems to be approximately realized in the {\it TESS} data ($g3$ is stronger in the first half of each $\sim$27 day-long sector, and weaker in the second), $g3$ also appears to be variable on shorter time-scales, which could be an artifact of its somewhat low amplitude or intrinsic variability in the amplitudes of one or both main frequencies, or it could reflect the presence of additional nearby frequencies with low amplitudes, analogous to the structure of $g2$. Fig.~\ref{fig:noise_spectrum} hints at the latter, where there seems to be structure above the Fourier `continuum' in the vicinity of $g3$ that extends beyond the two frequencies identified in Fig.~\ref{fig:multi_freqs} and Table~\ref{tbl:freqtbl}. Comparative conjectures concerning $g4$ are even more difficult, again owing to its low amplitude. Despite this, it is reasonable to claim that $g4$ is located at approximately three times the frequency of the strongest signal in $g2$ (and about 1.5 times the frequency of the strongest signal in $g3$), that $g4$ does not appreciably vary in frequency or amplitude from sector to sector, and that there does not seem to be any further structure besides the single detected peak above the noise level in Fig.~\ref{fig:noise_spectrum}.

\subsection{Stochastic variation, internal gravity waves, angular momentum transport, mode excitation, and Rossby waves} \label{sec:diss_stochastic}

Especially since internal gravity waves excited by core convection in differentially rotating massive stars are predicted to be efficient at transporting angular momentum from the core outwards \citep{Aerts2019,Lee2020}, this is an important topic to consider in Be stars, whose envelopes are somehow made to spin rapidly. If, as seems likely for many other OB stars, a spectrum of internal gravity waves is the origin of the bulk of the stochastic variability in $\gamma$\,Cas (Sec.~\ref{sec:stochastic}), this can aid in transporting angular momentum to the envelope, which would otherwise spin down as the star evolves and the envelope expands (and loses angular momentum to winds and mainly to the disk in the case of Be stars).  If angular momentum is transported to the outer layers of an already rapidly rotating star, then outbursts may be a way to remove this excess angular momentum, thus allowing the star to avoid super-critical rotation \citep{Krticka2011,Georgy2013,Rimulo2018,2020svos.conf...35B}.

A complication in the picture of the amplitude spectrum of $\gamma$\,Cas (and the majority of all early-type Be stars) is the presence of periodic signals in the standard g mode regime (namely, $g1$ and $g2$). However, this delineation is blurred by rapid rotation. 
With a spectral type of B0.5\,IVe, $\gamma$\,Cas is too hot for the $\kappa$ mechanism to drive g modes in the standard view of low frequency pulsation in SPB stars ($f \lesssim 4$ \cd), yet the bulk of its periodic signals lie in this regime. Global Rossby waves (r modes) are expected and observed in moderately to rapidly rotating stars \citep{VanReeth2016,Saio2018}, which tend to form frequency groups just below the rotation frequency (for $m = 1$) and/or just below twice the rotation frequency (for $m = 2$). In all the examples of \citet{Saio2018}, including one Be star, the $m = 1$ group is both predicted and observed to be significantly stronger than the $m = 2$ group. In the case of $\gamma$\, Cas, $g2$ is stronger than $g1$ (or even in the most conservative estimates they are of similar strength if the strongest signals in $g2$ are discounted), suggesting that $g1$ and $g2$ are not simply the manifestations of two groups of global r modes of $m = 1$ and $m = 2$. This bears resemblance to the Be star analyzed in \citet{Saio2018}, KIC 6954726, which has a first frequency group centered near 0.9~\cd that is consistent with r modes of $m = 1$, and a second group centered near 1.7 \cd (and of similar strength to the first group) that is inconsistent with $m = 2$ r modes (both in location and relative strength). As discussed in \citet{2016A&A...593A.106R}, this star was undergoing a quasi-continuous series of small outbursts at the time of the Kepler observations so that circumstellar activity likely contributes to aspects of the amplitude spectrum. r modes are necessarily retrograde in the co-rotating frame of the star, and thus do not positively contribute to the outward angular momentum transport (and in fact can slightly decrease the angular momentum in near-surface layers).  

It is possible that low-frequency g modes are excited by resonant coupling with internal gravity waves excited by core convection \citep{Lee2020}. The role of observable periodic signals excited in this way were explored for the Be star HD\,51452 (B0\,IVe), observed with the CoRoT satellite \citep{Neiner2012}. Like other early-type Be stars, the most prominent periodic signals are at frequencies between 0.5 -- 3 \cd, and are inconsistent with rotation and p mode oscillations. The proposed origin of these signals are stochastic gravito-inertial modes that are excited by core convection and modified by the Coriolis acceleration, in addition to possible Rossby waves (r modes) driven by the Coriolis acceleration. HD\,51452 is also mentioned in \citet{Saio2018}, where the authors suggest odd and even r modes of $m = 1$ as sufficient to explain the observed frequency groups without the need to invoke internal gravity waves driven by core convection as an excitation mechanism.

A recent study of the Be star HD\,49330 (B0.5\,IVe) by \citet{Neiner2020} demonstrates all the salient points of this section. HD\,49330 has similarities to $\gamma$\,Cas, being of the same spectral type and having frequencies in both the g and p mode regimes, although its disk is far weaker and less persistent compared to $\gamma$\,Cas. 
In an effort to understand the observed amplitude spectrum of HD\,49330 and its correlations with outbursts observed by CoRoT \citep{2009A&A...506...95H,2009A&A...506..103F}, \citet{Neiner2020} modeled various aspects of the system and found that stochastically excited internal gravity waves act to transport angular momentum to the surface layers, which then become destabilized and oscillate in stochastically driven g modes, whereupon outbursts can be triggered. \citet{Neiner2020} conclude that the $\kappa$ mechanism drives the observed p modes, but cannot drive the observed g modes. Instead, the g modes are excited by stochastic internal gravity waves.

The stochastic low frequency excess observed in the amplitude spectrum of $\gamma$\, Cas likely reflects the presence of internal gravity waves driven at the inner boundary of the radiative zone. This spectrum of internal gravity waves can excite g mode pulsation similar to the case of HD 49330, which can explain the presence of $g1$ and $g2$ in $\gamma$\, Cas. At the same time, r modes may (and probably do) exist which can also contribute to the observed frequency groups. Circumstellar activity may also cause photometric variability in the vicinity of $g1$ and $g2$. The $\kappa$ mechanism is likely the driving force behind the higher frequency signals in $g3$ and $g4$ (under the assumption they are p modes).  Given that the disk of $\gamma$\, Cas was growing at the time of the {\it TESS} observations, it is possible that all three of these processes (stochastically driven g modes, r modes, and circumstellar activity) are contributing to $g1$ and $g2$. Detailed modeling and spectroscopic identification of the line profile variability associated with these frequencies is needed to determine the physical processes responsible for, and their relative contribution to, the observed amplitude spectrum of $\gamma$\, Cas.

\subsection{Mass ejection and comparison to other early-type Be stars} \label{sec:comparison}

The signals recovered in the {\it TESS} data of $\gamma$\,Cas are perfectly ordinary compared to those found in large samples of classical Be stars \citep{2020svos.conf..137L, BalonaOzuyar2020}. There are a number of different types of variability seen in the brightness of Be stars, including frequency groups as well as individual isolated frequencies, stochastic variability, and longer time-scale (a few days and longer) changes in the mean brightness. In some Be stars monitored with the \textit{BRITE}-Constellation satellites, mean brightness variations by up to a few per cent and on time-scales of weeks to months have been observed at times when the amplitude of $g1$ modes was temporarily increased.  In some instances, the growth in amplitude appeared larger, and was more peaked in time, than would follow from the mere temporary agreement in phase (beating) of frequencies and can be due to non-linear amplification \citep{2016A&A...588A..56B, 2018A&A...610A..70B, 2020svos.conf...35B}.  The proposed explanation of the variable brightness \citep{Haubois2012, 2018MNRAS.479.2214G} is in terms of discrete mass-loss events that are driven by the non-linear amplitude superposition of several NRP modes and lead to an increased continuum emission by the ejecta. However, if the matter is ejected into the line of sight, the total brightness can be reduced.

In spite of the - in comparison to \textit{BRITE} and \textit{SMEI} \citep[\textit{e.g.}][]{Borre2020} - much shorter time baseline of less than a month (mostly one {\it TESS} sector), about one-quarter of early-type Be stars observed by {\it TESS} (B0 -- B3) show variability in their mean brightness on time-scales of a few days or longer and with amplitudes typically of a few percent \citep{2020svos.conf..137L,2020arXiv201013905L}. {\it TESS} data for four early-type Be stars are shown in Appendix~\ref{sec:massloss} to illustrate this point and serve as a comparison to $\gamma$\,Cas. In most, but not all, cases of days-long net brightening events, they are accompanied by increased amplitudes of one or more of the main frequency groups (mostly $g1$ or $g2$). By analogy to the much slower and often higher-amplitude events that are well understood to trace the build up of the circumstellar disk \citep[\textit{e.g.}][]{2016A&A...588A..56B, 2018A&A...610A..70B,LabadieB2018}, these more rapid changes in mean brightness are likely the result of varying amounts of material being lifted above the photosphere. This leads to the notion that, as the result of the temporary phase-coherent combination of multiple NRP modes, active Be stars can quasi-permanently (when averaged over months or years) eject matter that can be picked up by viscosity to build a disk.  The process may be stochastic as already discussed by \citet{2015MNRAS.450.3015K}.  

The {\it TESS} light curve of $\gamma$\,Cas does not clearly invite for a similar conclusion because the variations in mean brightness do not include clearly defined extrema (the mean brightness may also be affected by the signal and background extraction).  Accordingly, the time intervals of temporary in-phase superposition of multiple NRP modes that stand out as bright blobs in the wavelet transform but are also visible in the light curve (see bottom and top panels, respectively, of Fig.\,\ref{fig:wavelet}) do not clearly correlate with changes in the mean brightness.  
  
The lack of such a correlation neither invalidates the interpretation given above of the cases presented in Appendix~\ref{sec:massloss} nor is it expected that the correlation is seen in {\it TESS} data of a given Be star at all times or in all Be stars.  In fact, the saturation effect described in Sect.\,\ref{sec:spectra} even implies that in stars with a very dense inner disk small mass-ejection events may become photometrically very insignificant.  At the time of the {\it TESS} observations, the disk of $\gamma$\,Cas was clearly very strongly developed.  Furthermore, there can also be extended time intervals with significant mass loss but so small changes in its rate that the light curve is smooth over the duration of a single {\it TESS} sector \citep{LabadieB2017,Rimulo2018}.  Any superimposed discrete mass-loss events may also be so separated in time that a single {\it TESS} sector does not capture one of them.  Nevertheless, it is well possible that the longer term increase in brightness seen in Sectors 17 and 18 reflects enhanced mass loss rates over this time period. However, this is more ambiguous than the cases discussed in Appendix~\ref{sec:massloss}, where the evidence for brief mass ejection events is less hampered by a strong disk that has been growing for decades.

\subsection{Historical and new photometric frequencies} \label{sec:historical-freqs}
There is no clear indication in {\it TESS} Sectors 17, 18, or 24 of the $1.215811\pm0.000030$\,d period (0.82250(2)\,\cd frequency) found by \citet{HenrySmith} with an initial amplitude of $\sim$5-7\,ppt and confirmed by \citet{Borre2020} to have faded and eventually dropped below detectability.  In the combined {\it TESS} data, the closest signal appears at 0.8099(8)\cd and is the third strongest signal in $g1$ with an amplitude of 0.59\,ppt.  Formally, the 0.82\,\cd frequency has not, therefore, returned although, in view of the challenges discussed above of accurately and consistently measuring a frequency in a highly variable frequency group, the uncertainty may be larger than the numbers suggest.  \citet{2016AdSpR..58..782S} do not identify the physical property that is supposedly rotationally modulated with 0.82\,\cd, and they do not explain the phase coherence reported by \citet{SmithHenryVishniac} which is not an obvious property of weak, stochastically generated magnetic structures.  Therefore, there is no reason to pick a convenient frequency in $g1$ and call it the stellar rotation frequency.  In fact, from an investigation of {\it TESS} light curves of 15 $\gamma$\,Cas stars including $\gamma$\,Cas itself, \citet{2020MNRAS.498.3171N} do not report a single rotation frequency. 

As \citet{Borre2020} found, the 0.82\,\cd frequency seems to have given way to a newly developing variability with frequency 2.47936(2)\,\cd (which falls into $g2$) and time-dependent amplitudes between 1.5 and 4.5\,mmag (ranging between about 1 -- 3 mmag in the red filter of \textit{BRITE} and 3 -- 5 mmag in the blue filter). This amplitude difference considering the red and blue filters is expected for pulsation in hot stars, as photometric amplitudes are higher at shorter wavelengths. {\it TESS} found the latter of these signals still present in all sectors and at 2.4796(1)\,\cd in their combination.  In agreement with \citet{Borre2020} and the group behavior described above, the amplitude was variable between 2.6 and 3.7\,ppt but was at all times the largest of all individual signals (Table\,\ref{tbl:freqtbl}).  \citet{Borre2020} determined that the 0.82\,\cd and 2.48\,\cd frequencies are not 1:3 harmonics, having a conservatively estimated ratio of 3.0158(1).  The {\it TESS} value for the latter frequency is in agreement with this conclusion.  The 1.25\,\cd variability tentatively identified by \citet{Borre2020} is not confirmed by {\it TESS} (see Fig.\,\ref{fig:multi_freqs}).  The detection was declared tentative because the feature appeared surrounded by others in the power spectrum.  In this region, {\it TESS} now found group $g1$. The strongest signal in $g3$, 5.0559(5) \cd, seen by {\it TESS} is also apparent in the blue-filter \textit{BRITE} data at a similar amplitude \citep[about 1 ppt,][]{Borre2020}.  But, owing to its low SNR and proximity to an integer multiple of 1 \cd, the signal could not be unambiguously attributed to $\gamma$\,Cas. Since {\it TESS} does not suffer from daily aliases, this problem is avoided and the 5.06 \cd signal tentatively reported in \citet{Borre2020} is confirmed. 

Frequencies persistently present during the {\it TESS} observations were the strongest feature in $g2$ at 2.4796(1)\,\cd (and, more tentatively, at 2.3244(3) and 2.2663(4)\,\cd) and the two strongest ones in $g3$ at 5.0559(5) and 5.091(1)\,\cd as well as the seemingly isolated frequency at 7.571(1)\,\cd (see Table\,\ref{tbl1} and Fig.\,\ref{fig:multi_freqs}). Of these variabilities, the last one had the most constant amplitude. \citet{2020MNRAS.498.3171N} recently analyzed identical {\it TESS} data, but only for Sectors 17 and 18 and extracting the signal in a much less sophisticated way than described in Appendix\,\ref{sec:LC-extraction}.  They found frequencies of 2.480, 5.054, and 7.572\,\cd but do not give error estimates or amplitudes.

Frequency group $g1$, which in $\gamma$\,Cas includes the 0.82 and 1.25\,\cd frequencies, is, in other Be stars, near its lower limit often home to the {\v S}tefl frequencies (Sect.\,\ref{sec:gCas}) that are thought to trace activity in the innermost disk. In the latter, only in Sector 18 a feature at 1.115(4) \cd stands out above the remainder of $g1$ (Table\,\ref{tbl:freqtbl}).
The apparent lack of {\v S}tefl frequencies in photometry, however, does not exclude ongoing star-to-disk mass transfer.  Because {\v S}tefl frequencies may arise in newly ejected matter not yet in circularized orbits and/or not yet homogeneously distributed in azimuth, the indicated high inner-disk density could accelerate this process and reduce its amplitude. Frequent small events would further equalize the matter density in azimuth and circularize the orbits more quickly. The photometric signature of {\v S}tefl frequencies is also more likely to manifest in systems with higher inclination angles, whereas the intermediate inclination angle of $\gamma$\, Cas \citep[$\sim42^\circ$,][]{SteegCas} can be a hindrance to their detection. Furthermore, such circumstellar signals could exist among $g1$ and/or $g2$, but their identification as such would require spectroscopic confirmation.

\citet{1988PASP..100..233Y} reported periods between 7000 and 10000\,s for the traveling subfeatures in spectral lines (high-order NRP modes).  The corresponding frequency range from 8.6 to 12.3\,\cd in the {\it TESS} amplitude spectra is devoid of significant features, indicating long-term amplitude variations or cancellation of high-spatial-frequency signals in integral-light data.

\section{Conclusions}
\label{sec:conclusions}

The presence of frequency groups proves that $\gamma$\,Cas pulsates in low-order NRP modes because frequency groups have been found and attributed to low-order NRP in many Be stars, in $\beta$\,Cep and SPB stars, and even outside the B-star domain in $\gamma$\,Dor stars (cf.\ Introduction). 
Another commonality suggested by the {\it TESS} observations is the involvement of NRPs in several discrete mass-loss events within a month in some Be stars (Appendix~\ref{sec:massloss}).  At the time of the {\it TESS} observations, the strongly developed disk of $\gamma$\,Cas probably diluted the photometric evidence of individual events while the mean star-to-disk mass-transfer rate must have been high to let the disk grow even further.  Considering all other available observed properties, these findings reduce to a minimum the qualitative difference between $\gamma$\,Cas and the class of classical Be stars as a whole.  This concerns the properties of central stars and circumstellar disks alike as well as the physical processes governing them. \citet{2020MNRAS.498.3171N} recently arrived at a similar conclusion for a sample of 15 Be stars sharing the X-ray properties of $\gamma$\,Cas.

The establishment of low-order NRP also reinforces the spectroscopic identifications of higher-order NRP \citep{1988PASP..100..233Y} that have also been seen in other Be \citep{1983ApJ...275..661V, 1997ApJ...481..406K} and non-emission-line B stars \citep{SmithSpicaI}.  It is, therefore, very difficult to instead imagine undetectable magnetic fields producing azimuthally roughly equidistant super-photospheric cloudlets \citep{2016AdSpR..58..782S}. Even more challenging is it to anchor anything in a stellar atmosphere such that it does not show detectable phase wobble.  If the principle is that same objects require same explanations, $\gamma$\,Cas is to be explained by what explains other Be stars.  Since most other Be stars do not require an explanation of hard X-rays, because they do not emit them, and there is nothing in the standard toolbox for single Be stars suggesting that X-rays should arise anyway, the X-ray properties of $\gamma$\,Cas need to be explained by something that is not needed to understand the basics of Be stars.  A good candidate is the presence of helium-star companions \citep[][and references therein]{LangerBaade2020} that are tracers of an important formation channel of Be stars.

The picture conveyed by Appendix~\ref{sec:massloss} is that, in their equatorial regions, some Be stars are continuously bubbling like some hot springs on Earth.  The part made visible by {\it TESS} goes down to small events of up to a week (very few stellar rotations) in duration.  In addition, there can be sometimes cyclically repeating major events and a putative floor of activity powered by high-order NRPs and turbulence.  The relative contributions to the total mass loss to the disk are unknown.  In $\gamma$\,Cas, the star-to-disk mass-transfer rate seems to have increased in recent years because the profile of the emission in H\,$\alpha$ has slowly grown in EW and E/C and has changed from flat-topped to peaked, indicating an overall growth in disk mass and an increased density in the outer disk (Labadie-Bartz et al. in prep).

The 2.48\,\cd and 5.06\,\cd signals have probably persisted from 2015 \citep{Borre2020} through 2020 ({\it TESS}).
The 0.82\,\cd frequency found between 1997 and 2011 by \citet{HenrySmith} has not returned in 2019/2020.  It falls into $g1$ which exhibited amplitude variability over the {\it TESS} timeline.  Nevertheless, at $\sim6.5$\,mmag, the initial amplitude of the 0.82\,c/d variability was large.  This frequency seems too high for the {\v S}tefl frequency of $\gamma$\,Cas which is at the extreme high end of possible stellar rotational frequencies \citep{HenrySmith}.  Like 2.48\,\cd, it is likely just one of several frequencies in its parental group, and there is no reason to single it out for specific interpretative purposes.  The two magnetic models for Be stars zoom into just one frequency and, therefore, are not likely to capture the full picture.  Most notably, they do not seem to offer provisions for the (non-linear) coupling of several frequencies to initiate mass-loss events.  The case of $\pi$\,Aqr \citep{Naze2020} may show that $\gamma$\,Cas-like stars are not exempt from this.  0.82\,\cd and 2.48\,\cd are confirmed by \citet{Borre2020} to be not obviously numerically related despite being nearly at a 1:3 ratio.  

The lack of phase coherent variability in $g1$ seems in agreement with the nondetection of spectroscopic signatures of low-order NRP modes in $\gamma$\, Cas. On the other hand, the yearslong presence at high amplitude of the 0.82\,c/d frequency in $g1$ indicates that the nature of $g1$ may be more mixed than it appeared at the time of the {\it TESS} observations.  

The amplitude spectrum of $\gamma$\,Cas contains frequencies well above the conventional g mode domain.  This includes $g3$ near 5.1\,\cd and the seemingly solitary feature at 7.57\,c/d.  The latter should be outside the domain of rotationally split low-order g modes.  Because $\gamma$\,Cas is close to the hot limit of the $\beta$\,Cephei instability strip, it may be a hybrid $\beta$\,Cep/Be(SPB) pulsator \citep{2019A&A...632A..95M}.  This property is shared by about one-quarter to one-third of all Be stars in the $\beta$\,Cephei domain \citep{2020svos.conf..137L, BalonaOzuyar2020}. Like other early-type Be stars with pulsation in the traditional g mode domain, it remains to be explained what mechanism(s) act to drive these modes. Excitement by internal gravity waves is one attractive possibility \citep{Neiner2020}. Angular-momentum transport by gravity waves may aggravate the possible surface-angular-momentum crisis of Be stars \citep{Krticka2011,Rimulo2018}.  Mass loss driven by NRP would avoid this problem, so that NRP modes might couple such as to enable mass loss \citep{2020svos.conf...35B}. r modes may also contribute to these lower frequency groups.

\section*{Acknowledgements}

The authors are grateful for the comments from the anonymous referee, which have improved the manuscript. This paper includes data collected by the {\it TESS} mission, which are publicly available from the Mikulski Archive for Space Telescopes (MAST) and at the Space Telescope Science Institute (STScI). STScI is operated by the Association of Universities for Research in Astronomy, Inc., under NASA contract NAS 5–26555. Funding for the {\it TESS} mission is provided by NASA's Science Mission directorate.  This research has made use of NASA’s Astrophysics Data System \citep[ADS,][]{2000A&AS..143...41K}, the SIMBAD database \citep{2000A&AS..143....9W}, operated at CDS, Strasbourg, France and the BeSS database, operated at LESIA, Observatoire de Meudon, France: http://basebe.obspm.fr. 
J.L.-B. acknowledges support from FAPESP (grant 2017/23731-1). A.C.C. acknowledges support from CNPq (grant 307594/2015-7) and FAPESP (grant 2018/04055-8). 
Funding for the Stellar Astrophysics Centre is provided by The Danish National Research Foundation (Grant agreement no.: DNRF106).
This research made use of Astropy,\footnote{\url{http://www.astropy.org}} a community-developed core Python package for Astronomy \citep{astropy2013, astropy2018}. This work makes use of observations from the LCOGT network.

\section*{Data Availability}

All {\it TESS} data are available publicly. There are many ways to access the {\it TESS} data products, the most straightforward of which is via the MAST website (\url{https://mast.stsci.edu/}). The extracted light curves used in this work will be provided upon request to the corresponding author.
 



\bibliographystyle{mnras}
\bibliography{Main} 



\appendix

\section{Light curve extraction from {\it TESS} images} \label{sec:LC-extraction}

At an $I$-band magnitude around 2.4, $\gamma$\,Cas heavily saturates the {\it TESS} detectors. However, since primarily the core of the point spread function (PSF) and bloom columns are affected, valid brightness measurements can be extracted from the non-saturated pixels in the PSF wings (\textit{i.e.} halo photometry). Alternatively, a much larger aperture mask can be used if it includes all charge-containing pixels. The {\it TESS} CCDs are designed to conserve charge even for highly saturated objects. That is, excess charges from saturated pixels flow to neighboring unsaturated pixels along the column and can be collected and measured from them. Tests carried out prior to launch demonstrated this for stars at least as bright as 4th magnitude in the {\it TESS} band (according to the {\it TESS} instrument handbook v0.1\footnote{\url{https://archive.stsci.edu/missions/tess/doc/TESS_Instrument_Handbook_v0.1.pdf}}). Using an aperture mask that includes all pixels that collect charge from the incident flux of a very bright star (\textit{e.g.} $\gamma$\,Cas) is therefore a valid method for extracting a light curve provided that the brightness limit at which charge is no longer conserved does not significantly interfere with the astrophysical signals.

\begin{figure}
\centering
\includegraphics[width=0.45\textwidth]{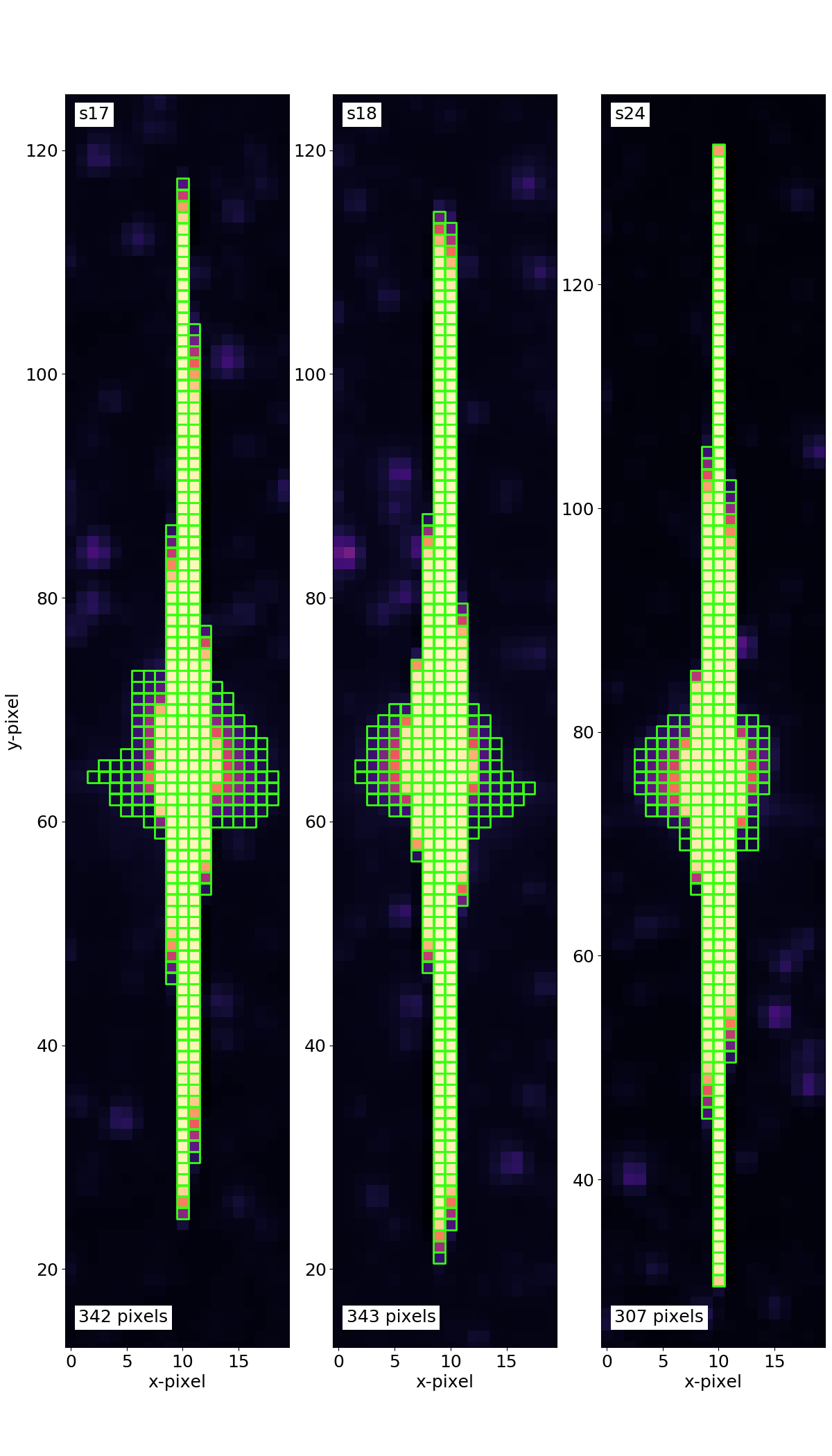}
\caption{{\it TESS} `postage stamp' showing the region of the sky around $\gamma$\,Cas for Sectors 17, 18, and 24. 
Pixels used to extract the light curve are marked in green. For the sake of increasing the contrast in this image, the plotted brightness of each pixel is proportional to the square root of its flux (averaged over all images in a given sector).}
\label{Fig:pixelmask}
\end{figure}

\begin{figure}
\centering
\includegraphics[width=0.48\textwidth]{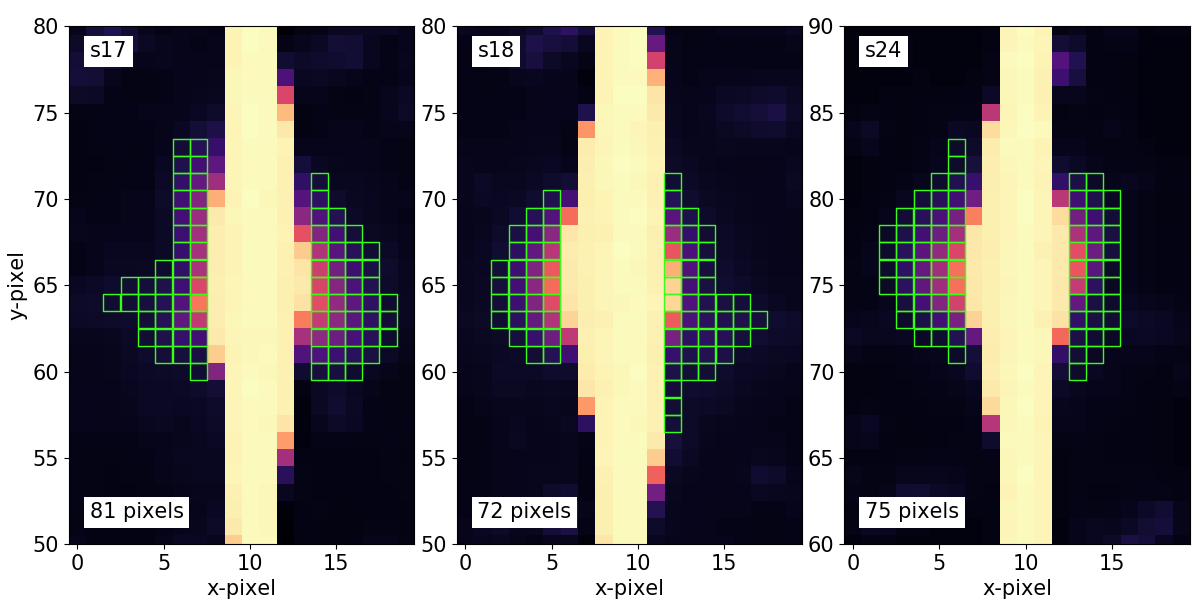}
\caption{{\it TESS} `postage stamp' in the same style as Fig.~\ref{Fig:pixelmask} showing the aperture mask used for halo photometry, where saturated pixels (and columns containing one or more saturated pixel) are not included.}
\label{Fig:pixelmask_halo}
\end{figure}

Light curves were extracted using both of these methods and compared. The aperture masks used are shown in Figs.~\ref{Fig:pixelmask} and~\ref{Fig:pixelmask_halo} for the methods using all charge-containing pixels and only including non-saturated halo pixels, respectively. 
For halo photometry, a threshold of 1000 $e^{-}s^{-1}$ was applied to the calibrated flux value of the postage stamp, averaged over all images in a sector and within a radius of 9 pixels from the central position of $\gamma$\,Cas on the detector, but all columns that included one or more saturated pixel were avoided. A higher threshold of 3000 $e^{-}s^{-1}$ was chosen for the aperture mask that includes saturated pixels. For reference, the mean sky background level is 175 $e^{-}s^{-1}$, 191 $e^{-}s^{-1}$, and 448 $e^{-}s^{-1}$ for sectors 17, 18, and 24, respectively.

\begin{figure*}
\centering
\includegraphics[width=0.99\textwidth]{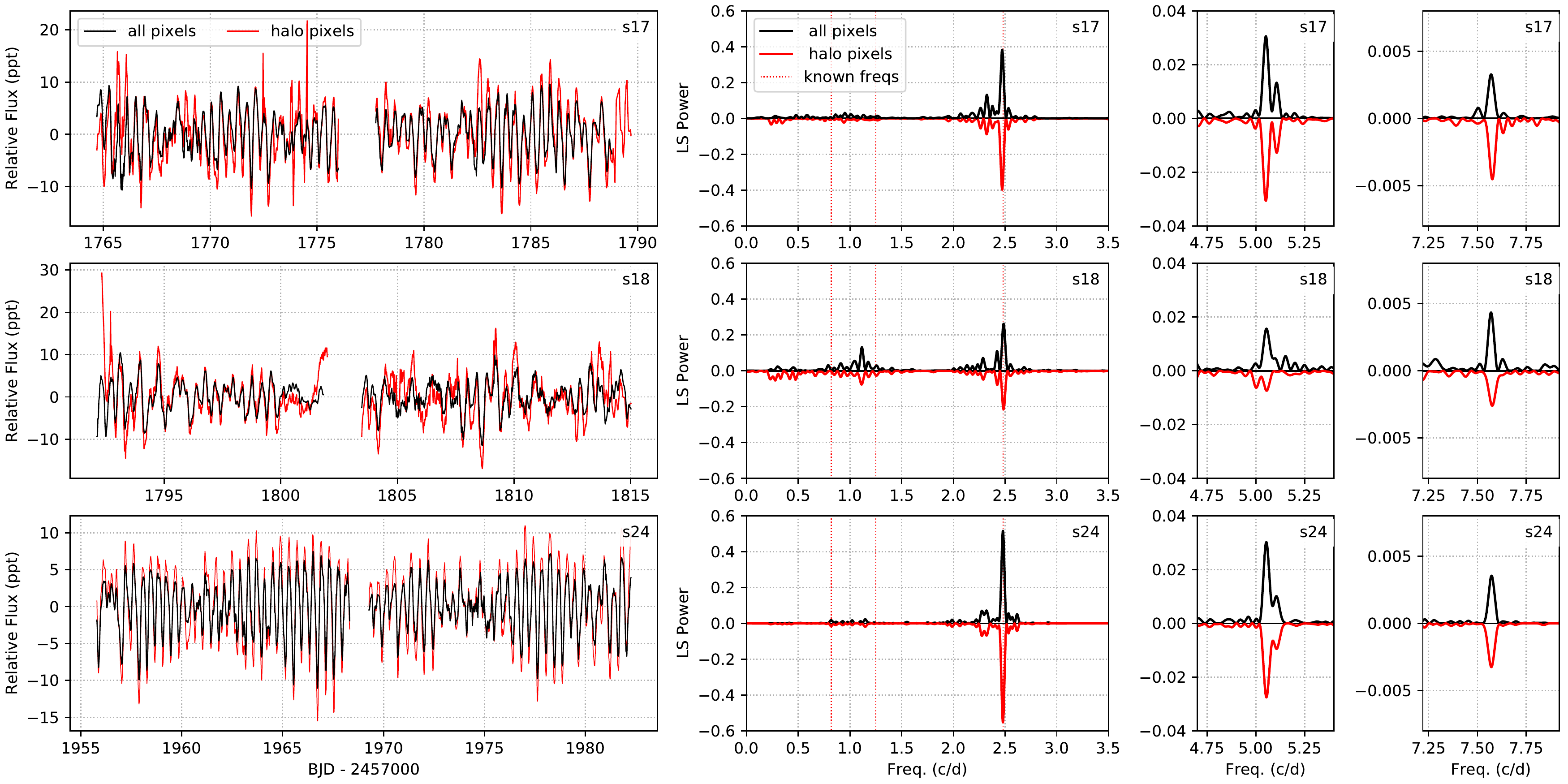}
\caption{\textit{Left:}  Comparison of the light curve for $\gamma$\,Cas extracted using all charge-containing pixels (black) and using only non-saturated halo pixels (red) for sectors 17, 18, and 24 after detrending. \textit{Right:} Lomb-Scargle periodogram for the light curve extracted using all charge-containing pixels (black) and using only non-saturated halo pixels (inverted, red) showing the main features. Owing to the different extraction apertures, there are slight differences in the systematic trends in the two versions of the light curves, requiring removal of data at slightly different times near the beginning and end of each half of each observing sector.}
\label{Fig:compare_halo_vs_sat}
\end{figure*}

The resulting light curves are compared in Fig.~\ref{Fig:compare_halo_vs_sat}, along with their respective power spectra, computed with the \textsc{timeseries.LombScargle} package \citep{VanderPlas2012,VanderPlas2015} of \textsc{Astropy} \citep{astropy2013, astropy2018}. The properties of these light curves are very similar, with the main differences being that halo photometry results in a lower SNR and an increase in apparent low frequency trends which are likely due to systematic effects. After removing linear and low frequency trends (with $f \leq 0.2$ \cd), the frequency spectra of the two versions of the data are essentially identical, in that the same signals are detected in both but with slightly higher amplitudes in the halo photometry (typically 10\% -- 40\% for the strongest periodic signal). 

The difference in amplitude between the two methods is most likely related to a varying PSF of $\gamma$\,Cas. The method using all relevant count-containing pixels is not affected by a variable PSF, since the PSF is sufficiently covered at both its widest and narrowest. However, since halo photometry only makes use of pixels in the non-saturated PSF wings, a variable PSF can change the fraction of the total flux from the source that falls into the halo pixels. In effect, the variability of $\gamma$\,Cas is `double counted' with halo photometry, leading to artificially enhanced amplitudes. Light variations are recorded due to the changing source brightness, but also due to a changing fraction of the ensquared PSF within the pixels selected. Put another way, when the source brightens, the PSF widens and an increased fraction of the source flux falls into the halo pixels. Therefore, the amplitudes associated with the aperture containing all count-containing pixels is deemed to be more reliable. The `breathing' of the PSF is readily seen in a movie of the images viewed over the duration of a given sector, provided a very high contrast ratio is used to view the images. Situations where halo photometry may provide more reliable results include when the bloom columns of a highly saturated object extend past the edges of the CCD (making it impossible to capture the total flux of the source), and when other relatively bright sources fall on or near the bloom columns (leading to contamination).

\citet{2020MNRAS.498.3171N} analyzed the same {\it TESS} data (Sectors 17 and 18 only, \textit{i.e.} without Sector 24).  However, they only applied aperture photometry to thumbnail images of 50$\times$50 pixels and thereby cut off the long blooming tails in the CCD columns.  Nevertheless they find basically the same frequencies as reported here which shows that the frequencies are not too sensitive to the signal-extraction strategy.  The amplitudes (and the signal-to-noise ratio) should be significantly lower because much of the vertically overflown signal was missed. Indeed, the amplitude of the strongest signal in $g2$ in this work is 3.67 ppt, and about 1.2 ppt in \citet{2020MNRAS.498.3171N}.

\section{Mass-loss indicators in {\it TESS} observations of other early-type Be stars}
\label{sec:massloss}

The following subsections show and discuss sample light curves for four early-type Be stars which 
illustrate variability patterns representative of the types of photometric signals seen in Be stars with {\it TESS} \citep{2020svos.conf..137L}. These light curves were extracted using methods similar to that for $\gamma$\,Cas (but without the complications of heavy saturation) and preserve the slow variability that illustrates that changes in the mean brightness and times of in-phase superposition of variabilities with several frequencies in $g1$ or $g2$ are correlated.  Because the time intervals concerned can be as long as $\sim$5\,d and several of them can occur in a single {\it TESS} sector of $\sim 27$\,d, the identification of such simultaneities is necessarily subjective.

Globally, there is no doubt that such a correlation exists, especially since qualitatively identical, probably cyclic correlations have been found on larger time-scales and at much higher amplitudes \citep{2017sbcs.conf..196B, 2018A&A...610A..70B}.  Only its prevalence cannot be quantified on the basis of the data used.  As explained in Sect.\,\ref{sec:discussion}, a change in mean magnitude can be the result of increased light scattering by additional matter ejected through the temporary (non-linear) in-phase superposition of several NRP modes. 

\subsection{HD\,58050 (HR\,2817, OT Gem, TIC\,14498757)}

\begin{figure}
\centering
\includegraphics[width=0.49\textwidth]{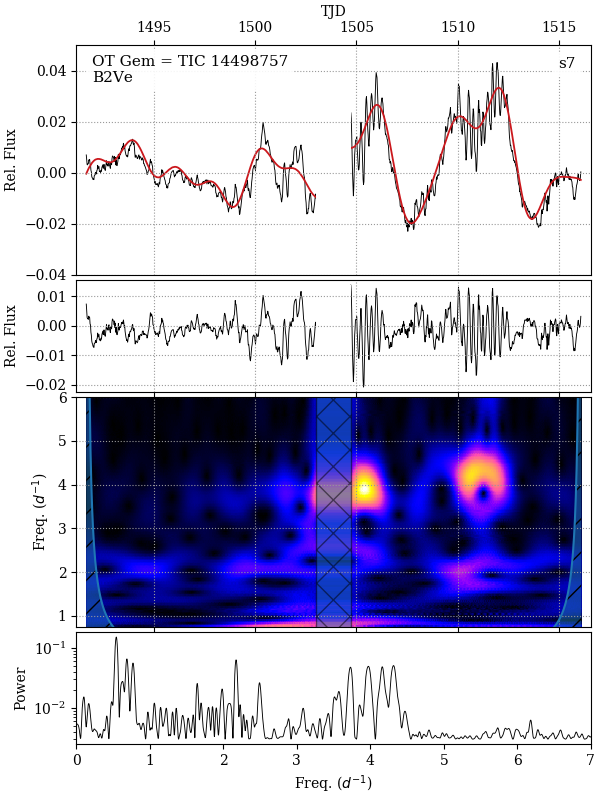}
\caption{\textit{Top:} Raw {\it TESS} light curve (after removing trends common to other pixels in the vicinity of the target) of OT Gem (TIC\,14498757) in black, with the red curve showing the low frequency component. \textit{Second:} {\it TESS} light curve after removing the low frequency component. \textit{Third:} Wavelet analysis showing the variability of the main frequency groups. \textit{Fourth:} Lomb-Scargle periodogram, computed after removing the low frequency component. }
\label{fig:14498757}
\end{figure}	

\citet{Bozic1999} distinguished photometrically quiet and active phases with peak-to-peak ranges of up to 0.4\,mag in $V$ and time-scales up to hundreds of days.  In observations obtained during the active phase they found cyclic variability with a repetition time of nearly 72\,d and a peak-to-peak amplitude of $\sim 0.15$\,mag.  The authors attributed this variability to binarity but did not have corroborating spectral evidence. 
Because of the link to enhanced photometric activity, a relation to the beating of NRP modes is equally plausible and supported by the triangular shape of the light curve, characteristic of short-lived mass ejection events \citep{Bernhard2018, LabadieB2018}

With a peak-to-peak range of $\sim 0.06$ mag, HD\,58050 is relatively active in regards to its secular brightening and fading events compared to other Be stars observed by {\it TESS}, but much less variable than during the active state described by \citet[][]{Bozic1999}, where the authors use a significantly longer time baseline. The {\it TESS} light curve (Fig.\,\ref{fig:14498757}) includes two rather strong local maxima during which the in-phase superposition of $g2$ frequencies is particularly clearly visible.

\subsection {HD\,71042 (TIC\,52929072)}

\begin{figure}
\centering
\includegraphics[width=0.49\textwidth]{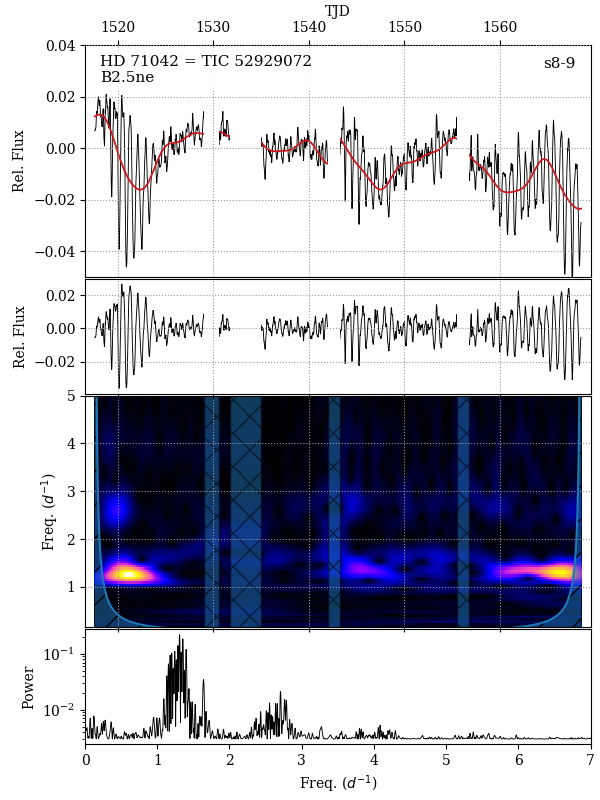}
\caption{Same as Fig.~\ref{fig:14498757}, but for HD\,71042 (TIC 52929072). }
\label{fig:52929072}
\end{figure}	

From observations in 3 of 5 consecutive annual seasons, \citet{LabadieB2018} classified HD\,71042 as Be star with outburst behavior and a full range of $\sim0.2$\,mag in $V$.  Contrary to the majority of Be stars, the outburst activity mainly consisted of dimmings rather than brightenings, suggesting that the line of sight passes through the disk.  There may be a 23.5\,d period with very unusual light curve. Observed in two consecutive {\it TESS} sectors, the light curve in Fig.\,\ref{fig:52929072} shows three time intervals when the mean magnitude dropped and subsequently ascended (the observations ended before a possible recovery after the third decline).  In all three intervals, the temporary in-phase superposition of $g1$ frequencies is prominent, and to a lesser extent those in $g2$.  This short-term variability, too, is consistent with a near-equator-on perspective.  However, there are no published spectra. The 23.5\,d period may also be present in the {\it TESS} data, although the short observational baseline makes this uncertain.

\subsection{HD\,144555 (TIC\,215511795)}

\begin{figure}
\centering
\includegraphics[width=0.49\textwidth]{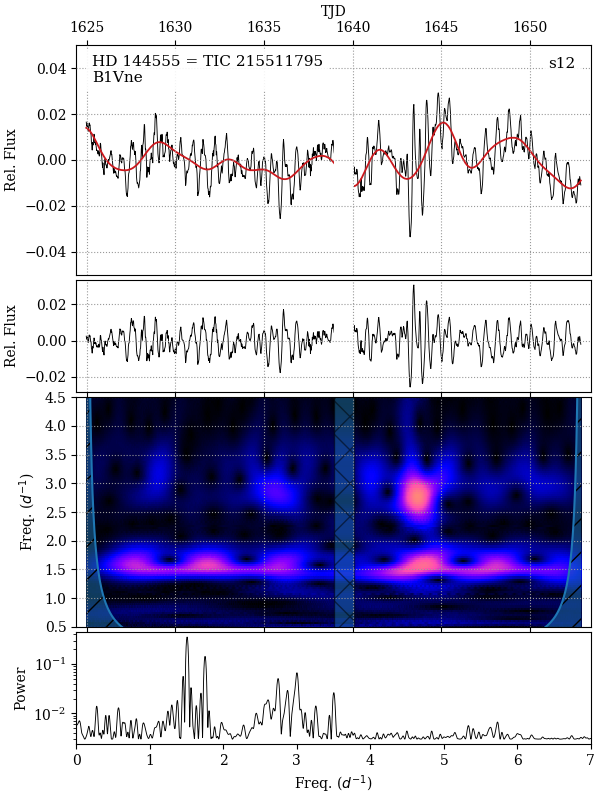}
\caption{Same as Fig.~\ref{fig:14498757}, but for HD\,144555 (TIC 215511795).}
\label{fig:215511795}
\end{figure}	

In Fig.\,\ref{fig:215511795}, five humps in mean magnitude appear in the light curve of HD\,144555.  On the ascending branch of all of them, variations with frequencies from $g1$ are temporarily co-phased. Around TJD = 1644 (TJD = BJD - 2457000), an additional phenomenon occurred in the form of a strong admixture of variability with $g2$ frequencies.  That multiple $g1$ as well as $g2$ frequencies are in phase both alone and together may be a coincidence.  But the probability of this happening in a single {\it TESS} sector of one month may be low.   Therefore, this phase could also be an indication of strong non-linearity as $g1$ and $g2$ are in a crudely harmonic relation. This phenomenon is seen in other Be stars observed by {\it TESS}. The difference between the two strongest signals in $g1$ is 0.244(1) \cd, consistent with the amplitude of $g1$ being modulated with a period of $\sim$4 days.

\subsection{HD\,156172 (TIC\,216875138)}

\begin{figure}
\centering
\includegraphics[width=0.49\textwidth]{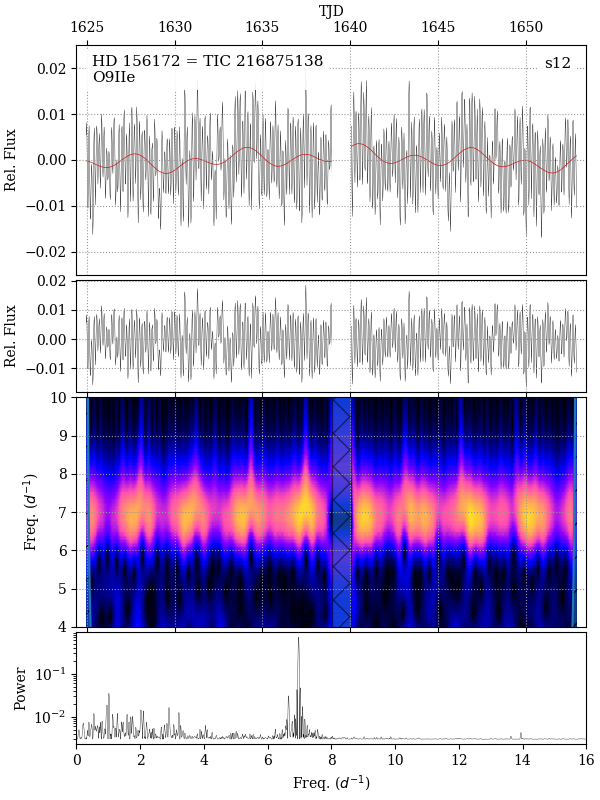}
\caption{Same as Fig.~\ref{fig:14498757}, but for HD\,156172 (TIC 216875138).}
\label{fig:216875138}
\end{figure}	

\begin{figure}
\centering
\includegraphics[width=0.49\textwidth]{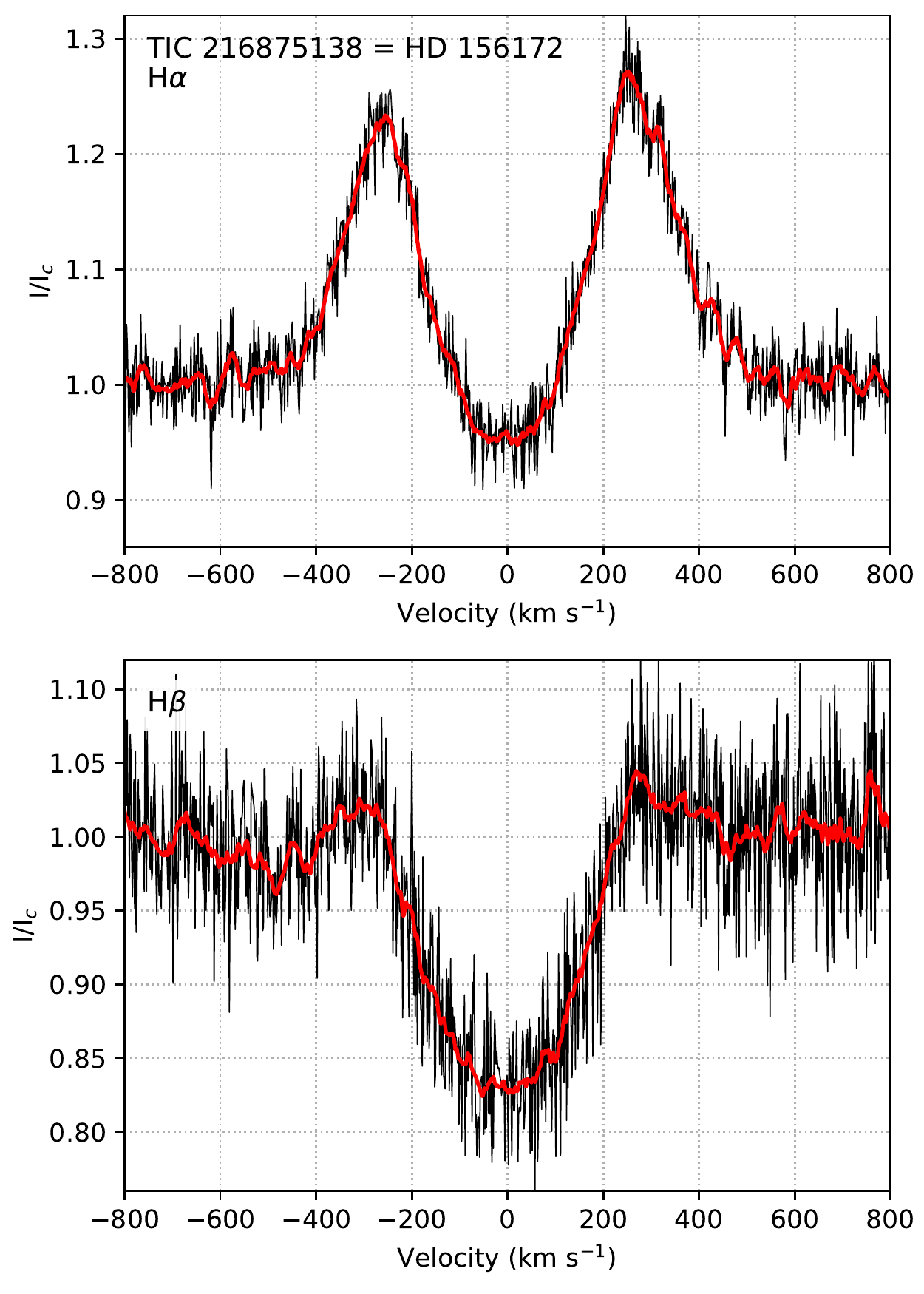}
\caption{The H\,$\alpha$ and H\,$\beta$ lines for HD\,156172 (TIC 216875138) from a single NRES spectrum obtained on 2020 July 01. Black (red) lines show un-binned (binned) data. }
\label{fig:bcep_Be}
\end{figure}

The photometric variability in Fig.\,\ref{fig:216875138} belongs to the most unusual ones described to date for any suspected classical Be star.  Firstly, the Lomb-Scargle amplitude spectrum of this late O-type star contains very little power below 4\,\cd, and frequency groups $g1$ and $g2$ are not discernable in a broad distribution of weak peaks.  The strongest features appear at the unusually high frequencies of 6.972(3) (amplitude $\sim$8 ppt) and 6.66(1)\,\cd (amplitude $\sim$2 ppt), respectively.  This has led to the classification of HD\,156172 as a $\beta$\,Cep star by \citet{Pigulski2018} who listed a single frequency of 6.97258(3)\,\cd and an amplitude of 12.4\,mmag in $V$. There is a much lower-amplitude pair of signals at frequencies of 13.947(8) and 13.63(1)\,\cd. The higher frequency of these is consistent with being the first harmonic of the strongest signal at 6.972(3)\,\cd, and the lower-frequency signal is equal (within errors) to the sum of the two frequencies near 7\,\cd. Secondly, the difference, 0.31(1)\,\cd, between these these two main frequencies appears as the beat frequency in the wavelet transform (Fig.\,\ref{fig:216875138}).  Since the beat frequency also exists in the amplitude spectrum (as the peak with the fourth highest amplitude, but not necessarily Lomb-Scargle power), there seems to be a non-linear mode interaction. Thirdly, the mean brightness of HD\,156172 varies with the same beat period and a semi-amplitude of 0.1 -- 0.2\%.  The amplitude is tiny but clearly detectable by plain visual inspection, and there is a well isolated peak in the unfiltered Lomb-Scargle spectrum at 0.312\,\cd. The maxima coincide with times of constructive interference of the said two frequencies (blobs in the wavelet transform).  

If these frequencies correspond to p modes, HD\,156172 might be one of the first Be stars in which indications of mass-loss due to p mode pulsations is found \citep[see also][]{Naze2020}.  Frequencies in the nominal p mode domain are not uncommon in Be stars \citep{2020svos.conf..137L, BalonaOzuyar2020}.  However, in HD\,156172, only the absence of the mostly much stronger $g1$ and $g2$ variabilities permitted the minute variation in mean brightness to be detected.  Generally, caution should be exercised in uncritically adopting a p mode identification merely on the basis of the observed frequencies because rapid rotation may greatly modify the frequencies in the inertial frame.  However, the absence of significant $g1$ and $g2$ frequencies as possible parent frequencies may make a p mode identification safer.  

Inspired by the above peculiarities, an echelle spectrum was obtained on 2020 July 01 with the NRES instrument attached to the 1-m telescope at South African Astronomical Observatory operated by LCOGT. The H\,$\alpha$ and H\,$\beta$ line are shown in Fig.~\ref{fig:bcep_Be}. This spectrum is entirely consistent with that of a classical Be star, and seemingly inconsistent with other types of objects that may show hydrogen emission (\textit{i.e.} interacting binaries, rapidly rotating OB stars with strong magnetic fields, supergiants, or Herbig Ae/Be stars). Other absorption lines are consistent with the spectral classification of a rapidly rotating late O or early B star. A more detailed investigation of this system is beyond the scope of this work, but is certainly warranted.


\bsp	
\label{lastpage}
\end{document}

%% file: freq_tbl_3.tex
\begin{table*}
 \caption{The frequencies above four times the red noise level are listed here, ordered by frequency group and amplitude. The SNR is calculated in the standard way in the \textsc{vartools} package by comparing the Lomb-Scargle power of a given peak to the mean value of the periodogram after applying iterative 5-sigma clipping. The logarithm of the formal false alarm probability (FAP) is also shown. The information in this table is from iterative pre-whitening of the light curve extracted from all count-containing pixels. Formal error estimates for the frequency determined according to \citet{Montgomery1999} are given in braces in units of the last significant digit. However, these should be understood as lower limits (see Section~\ref{sec:TSA}). Errors are not provided for amplitude or phase, because the variability of the signals makes such determinations difficult and somewhat arbitrary. } \label{tbl:freqtbl}
 \label{tbl1}
 \begin{tabular}{|cccccc|}
    \hline
        \textit{Group} &  \textit{Freq.} & \textit{Amp.} & \textit{Phase} & \textit{SNR} & \textit{log(FAP)}\\
                       &  \textit{(\cd)}    & \textit{(ppt)}  &         &          &       \\

\hline
\multicolumn{6}{|c|}{Sector 17}\\
\hline
$g1$ & 1.069(7) & 1.04 & -0.037 & 329.2 & -15.8\\
$g1$ & 1.008(7) & 1.00 & -0.114 & 344.2 & -13.0\\
$g1$ & 0.585(9) & 0.82 & -0.337 & 189.5 & -10.5\\
$g1$ & 0.90(1) & 0.74 & -0.189 & 152.5 & -10.2\\
$g1$ & 1.26(1) & 0.72 & -0.465 & 147.0 & -10.2\\
$g1$ & 0.80(1) & 0.64 & 0.100 & 109.2 & -9.4\\
\hline
$g2$ & 2.472(2) & 3.65 & 0.129 & 5224.6 & -85.8\\
$g2$ & 2.323(3) & 2.27 & -0.082 & 1934.1 & -52.5\\
$g2$ & 2.375(7) & 1.03 & 0.348 & 355.1 & -12.8\\
$g2$ & 2.194(9) & 0.85 & 0.324 & 226.4 & -10.8\\
$g2$ & 2.113(9) & 0.79 & -0.499 & 176.8 & -10.2\\
$g2$ & 2.44(1) & 0.76 & 0.263 & 161.8 & -9.9\\
$g2$ & 2.27(1) & 0.69 & 0.384 & 124.6 & -9.8\\
$g2$ & 1.95(1) & 0.60 & -0.339 & 106.8 & -10.0\\
$g2$ & 2.07(1) & 0.60 & 0.225 & 102.2 & -10.3\\
$g2$ & 2.63(1) & 0.59 & 0.083 & 99.6 & -8.8\\
$g2$ & 2.32(1) & 0.49 & -0.093 & 59.4 & -7.4\\
\hline
$g3$ & 5.051(7) & 1.04 & 0.124 & 325.1 & -12.1\\
$g3$ & 5.09(1) & 0.48 & 0.052 & 53.9 & -7.5\\
$g3$ & 4.86(3) & 0.24 & -0.449 & 13.5 & -4.3\\
$g3$ & 5.02(3) & 0.23 & 0.355 & 12.5 & -3.8\\
\hline
$g4$  & 7.56(2) & 0.37 & -0.028 & 26.6 & -5.6\\
\hline\hline
\multicolumn{6}{|c|}{Sector 18}\\
\hline
$g1$ & 1.115(4) & 1.85 & 0.125 & 1309.9 & -38.1\\
$g1$ & 0.967(8) & 1.00 & -0.179 & 274.1 & -13.5\\
$g1$ & 1.253(8) & 0.96 & 0.241 & 242.2 & -14.7\\
$g1$ & 0.788(8) & 0.96 & 0.487 & 239.7 & -16.2\\
$g1$ & 0.85(1) & 0.75 & 0.468 & 126.6 & -10.9\\
$g1$ & 1.03(1) & 0.72 & 0.136 & 108.4 & -11.3\\
\hline
$g2$ & 2.485(3) & 2.56 & -0.326 & 2176.9 & -54.9\\
$g2$ & 2.286(6) & 1.24 & 0.066 & 498.6 & -19.6\\
$g2$ & 2.417(8) & 0.98 & 0.117 & 260.5 & -14.2\\
$g2$ & 2.18(1) & 0.80 & 0.002 & 160.8 & -11.6\\
$g2$ & 2.06(1) & 0.64 & 0.009 & 99.6 & -9.4\\
$g2$ & 2.68(1) & 0.57 & 0.445 & 72.3 & -8.6\\
$g2$ & 2.13(1) & 0.50 & 0.187 & 45.3 & -7.5\\
$g2$ & 2.50(1) & 0.47 & 0.366 & 37.1 & -7.9\\
\hline
$g3$ & 5.05(1) & 0.64 & 0.235 & 100.0 & -10.0\\
$g3$ & 5.14(2) & 0.33 & -0.175 & 17.2 & -4.7\\
$g3$ & 5.09(3) & 0.25 & -0.477 & 12.1 & -3.2\\
\hline
$g4$  & 7.56(2) & 0.31 & -0.381 & 16.7 & -4.5\\
  \hline
 \end{tabular}
 \begin{tabular}{|cccccc|}
 \hline
        \textit{Group} &  \textit{Freq.} & \textit{Amp.} & \textit{Phase} & \textit{SNR} & \textit{log(FAP)}\\
                       &  \textit{(\cd)}    & \textit{(ppt)}  &         &          &       \\
\hline
\multicolumn{6}{|c|}{Sector 24}\\
\hline
$g1$ & 1.148(8) & 0.80 & -0.006 & 354.0 & -19.4\\
$g1$ & 0.816(9) & 0.69 & -0.085 & 265.3 & -16.9\\
\hline
$g2$ & 2.479(1) & 3.69 & 0.176 & 7640.2 & -134.1\\
$g2$ & 2.314(4) & 1.42 & 0.305 & 1134.1 & -38.9\\
$g2$ & 2.277(5) & 1.25 & 0.024 & 920.6 & -35.3\\
$g2$ & 2.411(6) & 0.94 & -0.067 & 515.3 & -22.2\\
$g2$ & 2.619(9) & 0.73 & 0.329 & 285.2 & -17.3\\
$g2$ & 2.57(1) & 0.61 & -0.135 & 194.8 & -15.0\\
$g2$ & 2.53(1) & 0.60 & 0.252 & 195.1 & -13.2\\
$g2$ & 1.98(1) & 0.58 & 0.000 & 163.5 & -14.1\\
$g2$ & 2.05(1) & 0.58 & -0.344 & 166.8 & -15.9\\
\hline
$g3$ & 5.053(7) & 0.91 & -0.133 & 476.7 & -23.1\\
$g3$ & 5.09(1) & 0.44 & 0.283 & 89.9 & -10.3\\
$g3$ & 5.13(2) & 0.24 & -0.385 & 24.5 & -5.4\\
$g3$ & 5.19(3) & 0.19 & 0.193 & 15.6 & -4.1\\
\hline
$g4$  & 7.57(2) & 0.29 & -0.343 & 36.1 & -5.9\\
\hline\hline
\multicolumn{6}{|c|}{All Sectors}\\
\hline
$g1$ & 1.1346(6) & 0.71 & 0.245 & 572.6 & -21.8\\
$g1$ & 1.0206(7) & 0.62 & 0.133 & 709.6 & -25.5\\
$g1$ & 0.8099(8) & 0.59 & -0.350 & 373.0 & -16.1\\
\hline
$g2$ & 2.4796(1) & 3.67 & -0.218 & 13244.3 & -256.9\\
$g2$ & 2.3244(3) & 1.30 & -0.400 & 2356.9 & -68.0\\
$g2$ & 2.2663(4) & 1.01 & 0.442 & 1289.5 & -40.7\\
$g2$ & 2.4161(5) & 0.95 & -0.483 & 487.1 & -19.0\\
$g2$ & 2.2902(7) & 0.69 & 0.255 & 476.6 & -19.9\\
$g2$ & 2.5421(8) & 0.56 & -0.085 & 348.7 & -16.1\\
$g2$ & 2.6331(8) & 0.55 & 0.034 & 297.8 & -14.8\\
$g2$ & 2.1187(9) & 0.53 & 0.041 & 390.1 & -16.1\\
$g2$ & 1.9452(9) & 0.52 & -0.142 & 297.6 & -14.2\\
$g2$ & 2.3607(9) & 0.50 & -0.257 & 577.8 & -21.6\\
\hline
$g3$ & 5.0559(5) & 0.90 & -0.382 & 934.0 & -28.9\\
$g3$ & 5.091(1) & 0.45 & 0.319 & 269.8 & -14.1\\
\hline
$g4$  & 7.571(1) & 0.30 & -0.442 & 103.4 & -8.1\\
  \hline
\multicolumn{6}{c}{ }\\
\multicolumn{6}{c}{ }\\
\multicolumn{6}{c}{ }\\
\multicolumn{6}{c}{ }\\
\multicolumn{6}{c}{ }\\
\multicolumn{6}{c}{ }\\
\multicolumn{6}{c}{ }\\
\multicolumn{6}{c}{ }\\
 \end{tabular}
 \begin{flushleft}
  \end{flushleft}  
\end{table*}